\documentclass{JHEP3}
\usepackage{diagrams}
\usepackage{oldgerm}

\newcommand{\Ibb}[1]{ {\rm I\ifmmode\mkern -3.6mu\else\kern -.2em\fi#1}}
\newcommand{\ibb}[1]{\leavevmode\hbox{\kern.3em\vrule
     height 1.2ex depth -.3ex width .2pt\kern-.3em\rm#1}}
\newcommand{\Cl}{{\ibb C}}           
\newcommand{\Rl}{{\Ibb R}}           
\newcommand{\Zl}{\mathbb{Z}}
\newcommand{\Nl}{{\Ibb N}}

\newcommand{\Om}{\Omega}
\newcommand{\om}{\omega}

\newcommand{\Hil}{\mathcal{H}}
\newcommand{\VV}{\mathcal{V}}
\newcommand{\We}{\mathscr{W}}

\newcommand{\E}{\mathcal{E}}
\newcommand{\U}{\mathcal{U}}
\newcommand{\DD}{\mathcal{D}} 
 
\newcommand{\W}{\mathcal{W}}    

\newcommand{\Ss}{\mathscr{S}}   
\newcommand{\uSs}{\underline{\mathscr{S}}}   
\newcommand{\Ssd}{\mathscr{S}(\Rl^4)}   
\newcommand{\Ssdn}{\mathscr{S}(\Rl^{n})}   
\newcommand{\SsKd}{\mathscr{S}_K}   
\newcommand{\SsKdn}{\mathscr{S}_K^{n}}   
\newcommand{\SsKdm}{\mathscr{S}_K^{m}}   
\newcommand{\Ssdm}{\mathscr{S}(\Rl^{m})}   
\newcommand{\OO}{O}   
\newcommand{\PGc}{\tilde{\mathcal{P}}_+^\uparrow}   
\newcommand{\LGpo}{\mathcal{L}_+^\uparrow}

\newcommand{\LG}{\mathcal{L}}
\newcommand{\LGp}{\mathcal{L}_+}

\newcommand{\SL}{\text{SL}(2,\Cl)}

\newcommand{\Pol}{\mathscr{P}}

\newcommand{\xh}{\hat{x}}

\def\bx{{\mbox{\boldmath{$x$}}}}

\def\bof{{\mbox{\boldmath{$f$}}}}
\def\bog{{\mbox{\boldmath{$g$}}}}

\def\boh{{\mbox{\boldmath{$h$}}}}
\def\bk{{\mbox{\boldmath{$k$}}}}

\def\bm{{\mbox{\boldmath{$m$}}}}

\newcommand{\sbm}{{\mbox{\scriptsize \boldmath $m$}}}
\newcommand{\sbk}{{\mbox{\scriptsize \boldmath $k$}}}
\newcommand{\sbt}{{\mbox{\scriptsize \boldmath $t$}}}
\newcommand{\sbl}{{\mbox{\scriptsize \boldmath $l$}}}

\newcommand{\sbr}{{\mbox{\scriptsize \boldmath $r$}}}

\newcommand{\supp}{\mathrm{supp}\,}

\newcommand{\te}{\theta}

\newcommand{\la}{\lambda}
\newcommand{\La}{\Lambda}
\newcommand{\eps}{\varepsilon}

\newcommand{\kae}{\kappa_{\rm e}}
\newcommand{\kam}{\kappa_{\rm m}}

\newcommand{\omti}{\tilde{\omega}}
\newcommand{\fti}{\widetilde{f}}
\newcommand{\hti}{\widetilde{h}}

\newcommand{\Psiti}{\tilde{\Psi}}

\newcommand{\Uti}{\tilde{\cal U}}

\newcommand{\phiti}{\widetilde{\phi}}
\newcommand{\chiti}{\widetilde{\chi}}

\newcommand{\gti}{\widetilde{g}}

\newcommand{\fbar}{\overline{f}}

\newcommand{\kbar}{\overline{k}}

\newcommand{\phiot}{\phi^\otimes}

\newcommand{\phite}{\phi^\te}
\newcommand{\phitite}{\tilde{\phi}^\te}
\newcommand{\ot}{\otimes}

\newcommand{\otte}{\otimes_\theta}

\usepackage{amsmath}
\usepackage{amsfonts,amssymb}
\usepackage{mathrsfs}

\newtheorem{theorem}{Theorem}[section]
\newtheorem{proposition}[theorem]{Proposition}
\newtheorem{lemma}[theorem]{Lemma}

\newtheorem{definition}[theorem]{Definition}

\title{Noncommutative Deformations of Wightman Quantum Field Theories}
\author{Harald Grosse\\ Faculty of Physics, University of Vienna,\\ Boltzmanngasse 5, A-1090 Vienna, Austria\\ E-mail: \email{harald.grosse@univie.ac.at}}
\author{Gandalf Lechner\\ Faculty of Physics, University of Vienna,\\ Boltzmanngasse 5, A-1090 Vienna, Austria\\ E-mail: \email{gandalf.lechner@univie.ac.at}}

\abstract{
Quantum field theories on noncommutative Minkowski space are studied in a model-independent setting by treating the noncommutativity as a deformation of quantum field theories on commut	ative space. Starting from an arbitrary Wightman theory, we consider special vacuum representations of its Weyl-Wigner deformed counterpart. In such representations, the effect of the noncommutativity on the basic structures of Wightman theory, in particular the covariance, locality and regularity properties of the fields, the structure of the Wightman functions, and the commutative limit, is analyzed. Despite the nonlocal structure introduced by the noncommutativity, the deformed quantum fields can still be localized in certain wedge-shaped regions, and may therefore be used to compute noncommutative corrections to two-particle S-matrix elements.
}


\begin{document}

\section{Introduction}

Models of quantum field theories on deformed, noncommutative spaces have been under intensive investigation in the last years \cite{DoplicherFredenhagenRoberts:1995, Szabo:2003, GrosseWulkenhaar:2005,  BahnsDoplicherFredenhagenPiacitelli:2005, Rivasseau:2007, GrosseLechner:2007, AkoforBalachandranJoseph:2008}. A main motivation for studying such spaces is the fact that their spatial and temporal coordinates satisfy those uncertainty relations which are suggested by the uncertainty principle and classical gravity \cite{DoplicherFredenhagenRoberts:1995}. Quantum field theory on noncommutative spacetime therefore provides an intermediate step towards a full quantum mechanical treatment of gravity as required for Planck scale physics.

The construction of models on such deformed spaces faces however new difficulties, for example the nonlocal features caused by the noncommutative structure of the underlying space. In most approaches, noncommutative spaces are taken as a motivation for introducing modified effective Lagrangeans on commutative spaces. The corresponding field theories are then studied with the methods of perturbative renormalization, either in a Lorentzian or Euclidean setting \cite{BahnsDoplicherFredenhagenPiacitelli:2005, GrosseVignesTourneret:2008, GurauMagnenRivasseauTanasa:2008}, which sometimes are better behaved than in the commutative case \cite{GrosseWulkenhaar:2004, DisertoriRivasseau:2007, DisertoriGurauMagnenRivasseau:2007}.

Besides these constructions of specific models, there also exist model-independent proposals about the formulation of quantum field theories on noncommutative spaces \cite{AlvarezGaumeVazquezMozo:2003, ChaichianMnatsakanovaNishijimaTureanuVernov:2004, Soloviev:2008}. In this paper, we develop a particular model-independent approach and consider the noncommutativity as a deformation of a quantum field theory on commutative Minkowski space. Our starting point is an arbitrary theory of Wightman quantum fields $\phi_1,...,\phi_K$, about which we assume the usual covariance, locality and regularity properties, but make no assumptions as far as a Lagrangean formulation or the interaction is concerned. We then propose a deformed, noncommutative version of this theory, and study its properties.

As a simple and well studied example of a noncommutative space, we work on the so-called noncommutative Minkowski space. In the formulation given by Doplicher, Fredenhagen and Roberts \cite{DoplicherFredenhagenRoberts:1995}, it is modelled as a $C^*$-algebra generated by four selfadjoint coordinate operators $X_0,...,X_3$ and an identity $1$, satisfying the ``quantum conditions''
\begin{align}\label{NCMS}
 [X_\mu,X_\nu]=:i\,Q_{\mu\nu}&\,,\qquad [X_\mu, Q_{\nu\kappa}]=0\,,\\
 Q_{\mu\nu}Q^{\mu\nu}=2(\kae^2-\kam^2)\cdot 1\,&,\qquad \eps_{\mu\nu\la\rho}Q^{\mu\nu}Q^{\la\rho}=-8\kae\kam\cdot 1\,,
\end{align}
with some constants $\kae,\kam\in\Rl$, measuring the strength of noncommutative effects. Generalizing the well-known Weyl-Wigner correspondence between functions on commutative and noncommutative Minkowski space (see, e.g. \cite{Szabo:2003}), we consider the Weyl-Wigner deformed fields \cite{DoplicherFredenhagenRoberts:1995}
\begin{align}
X \longmapsto \int d^4p\;  e^{ip_\mu X^\mu}\ot \phiti_k(p) \,.
\end{align}
This formal assignment, well known from free field theories in the noncommutative setting \cite{DoplicherFredenhagenRoberts:1995}, is here used to define the polynomial algebra of general deformed quantum fields. To make contact with field theory on the Moyal plane \cite{Szabo:2003}, where the commutators $Q_{\mu\nu}$ are realized as multiples of the identity, we then consider special vacuum states on this field algebra, which correspond to fixing a value $\te$ in the joint spectrum $\Sigma$ of the commutators $Q_{\mu\nu}$. In the corresponding vacuum representations, we find a family of deformed quantum fields $\phite_k$, which coincide with the ones recently proposed by Soloviev \cite{Soloviev:2008}.

The main characteristics of the deformed theories governed by the fields $\phite_k$ can be summarized as follows: The continuity and domain properties of the Wightman fields are stable under the deformation, and in the commutative limit $\te\to0$, the deformed fields converge strongly to the undeformed ones. The deformed models are Poincar\'e covariant, but in general, Lorentz transformations link fields with different spectral values $\te\in\Sigma$, and we therefore consider an infinite family of different field operators. Although the fields $\phite_k$ are not local, we find that they are not completely delocalized either: Each field operator $\phite_k(f)$ can be localized in a certain wedge-shaped region of Minkowski space in a manner which is consistent with covariance and causality. This weak form of locality is strong enough to allow for the computation of two-particle S-matrix elements \cite{BorchersBuchholzSchroer:2001,GrosseLechner:2007,BuchholzSummers:2008}.
\\\\
These findings generalize our previous analysis of a free, scalar quantum field on noncommutative Minkowski space \cite{GrosseLechner:2007}, which was subsequently generalized by Buchholz and Summers to arbitrary models \cite{BuchholzSummers:2008} in the framework of algebraic quantum field theory \cite{Haag:1996}. From the point of view of deformations of observable algebras, we show in this paper how their general deformation theory looks like in a concrete Wightman setting, and how it is related to vacuum representations of Weyl-Wigner deformed fields. Moreover, we find that the deformation induced by the noncommutative space \eqref{NCMS} is only a single example of a large class of deformations, and mention other examples.
\\\\
This article is organized as follows: In Section \ref{sec:defqfs}, we summarize our assumptions on the underlying undeformed quantum field theory, and describe the form of noncommutative Minkowski space which we use. We then formulate the algebra of Weyl-Wigner deformed fields, introduce a class of vacuum states on it and consider the corresponding vacuum representations.

It turns out that the deformation of the Wightman theories amounts to a deformation of the underlying Borchers-Uhlmann tensor algebra of test functions, which is here endowed with a twisted (Moyal) tensor product instead of the usual tensor product. In Section \ref{Sec:MTP} we study various features of this Moyal tensor product, which are then used in Section \ref{Sec:te-fields} to derive the above mentioned properties of the deformed quantum fields.

Our conclusions and some comments on possible future developments are presented in Section \ref{sec:conclusion}.
\\\\
The following notations and conventions will be used throughout this paper. All our considerations take place on four-dimensional Minkowski space, the generalization to arbitrary dimensions $d\geq2$ being straightforward. We define the Minkowski metric as $\eta=\text{diag}(+1,-1,-1,-1)$, i.e. the inner product is $x\cdot y :=(x,\eta y) = x_0y_0-\sum_{k=1}^{3}x_ky_k$, where $(\,.\,,\,.\,)$ denotes the positive definite Euclidean scalar product. Also the Fourier transform is defined using the Minkowski product, $\fti(p) := (2\pi)^{-2}\int d^4x\,f(x)\,e^{-ip\cdot x}$. We will employ the notations
\begin{align}\label{te-conv}
 p \te q := (p,\eta \te \eta q) = p_\mu\te^{\mu\nu}q_\nu
\,,\qquad
(\te p)_\mu := \te_{\mu\nu} p^\nu\,,
\end{align}
and denote the space of all real, antisymmetric $(4\times 4)$-matrices by $\Rl^{4\times 4}_-$. Finally, a dash on a subset $\OO\subset\Rl^4$ is used to denote the causal complement of that region, $\OO'=\{x\in\Rl^4\,:\,(x-y)^2<0\;\forall y\in\OO\}$.

\section{Vacuum representations of deformed quantum fields}\label{sec:defqfs}

\subsection{Assumptions on the undeformed field theory}

The starting point of our investigations of deformed quantum field theories is an undeformed, usual theory on commutative Minkowski space $\Rl^4$, described in the Wightman framework \cite{BogolubovLogunovTodorov:1975, StreaterWightman:1964}. In this section, we collect our corresponding notations and conventions, which are by and large standard.

The theory is formulated on a separable Hilbert space $\Hil$, on which the relativistic symmetries act via a strongly continuous, (anti-) unitary representation $U$ of the universal covering group $\PGc$ of the identity component of the Poincar\'e group, $\PGc=\Rl^4\ltimes \SL$. We denote the covering homomorphism between $\SL$ and the identity component $\LGpo$ of the Lorentz group by $A\mapsto\La(A)$.

Since we are interested in vacuum representations, we require positivity of the energy in all Lorentz frames, i.e. the joint spectrum of the generators $P_\mu$ of the translation groups $U(y_\mu,1)$ lies in the closed forward lightcone $V_+:=\{p\in\Rl^4\,:\,p^2\geq0,\,p_0\geq0\}$. Furthermore, there exists a $U$-invariant unit vector $\Om\in\Hil$, representing the vacuum state.

We allow for finitely many arbitrary scalar, tensor or spinor fields and denote the components of all these fields by $\phi_1,...,\phi_K$, $K<\infty$. They constitute the operator-valued distribution $\phi(f):=\sum_{k=1}^K\phi_k(f_k)$, depending linearly on multi component test functions $f=(f_1,...,f_K)\in\Ss_K:=\Ssd^{\oplus K}$.

The operators $\phi(f)$, $f\in\Ss_K$, and their adjoints contain a common, stable, $U$-invariant, dense subspace $\DD\subset\Hil$ including $\Om$ in their domains such that $f\mapsto\langle\Psi,\phi(f)\Psi'\rangle$ is a $K$-component tempered distribution if $\Psi,\Psi'\in\DD$. More specifically, we will consider the fields as operators on the domain they generate from the vacuum, i.e. on
\begin{align}
 \DD &:= \text{span}\{\Psi^n(f^n)\,:\,f^n\in\SsKdn\,,n\in\Nl_0\}
\label{def:D}
	\,,\\
 \Psi^n(f_1\ot...\ot f_n) &:= \phi(f_1)\cdots\phi(f_n)\Om\,,\qquad f_1,...,f_n\in\SsKd\,,\label{def:Psin}
\end{align}
and assume that $\DD$ lies dense in $\Hil$. By application of the nuclear theorem, the $\Psi^n$ \eqref{def:Psin} an be extended to $\Hil$-valued tempered distributions on $\SsKdn$ \cite{Jost:1965}, i.e. we have a collection of linear, continuous maps $\SsKdn\ni f^n\mapsto\Psi^n(f^n)\in\Hil$.

It will be convenient to include the adjoints of the fields in the set $\{\phi_1,...,\phi_K\}$, and consider also them as being defined on the domain $\DD$. We write ${\phi_k(f)}^*|_\DD=\phi_{\overline{k}}(\fbar)$, $\overline{k}\in\{1,...,K\}$.

Depending on the transformation behavior of the fields $\phi_k$, there exists some $K$-dimensional representation $D$ of $\SL$ such that
\begin{align}
U(y,A)\phi(f)U(y,A)^{-1}
&=
\phi(f_{(y,A)})\,,\qquad f\in\SsKd\,,
\\
f_{(y,A)}(x)_k &:= \sum_{l=1}^K D(A^{-1})_{lk}f(\La(A)^{-1}(x-y))_l
\label{PGonBUA}
\,.
\end{align}

To describe the commutation relations of the fields, we assume that the index set $\{1,...,K\}=I_B\cup I_F$ is the disjoint union of a set $I_B$ of ``Bose indices'' and a set $I_F$ of ``Fermi indices''. The corresponding fields commute or anticommute at spacelike separation, i.e. with $\Psi\in\DD$, $f,g\in\Ssd$, 
\begin{align}\label{locality}
 [\phi_k(f),\,\phi_l(g)]_\pm\Psi
=
0\qquad\text{if}\;\;(\supp f)\subset(\supp g)'\;.
\end{align}
Here the sign in $[\phi_k,\,\phi_l]_\pm=\phi_k\phi_l\pm\phi_l\phi_k$ is ``$+$'' if $k,l\in I_F$ and ``$-$'' otherwise. Note that according to our conventions, these commutation relations also involve the adjoint fields $\phi_k(f)^*$.
\\
\\
It will often be convenient to consider the Borchers-Uhlmann algebra \cite{Borchers:1962, LassnerUhlmann:1968}, i.e. the tensor algebra $\uSs$ over $\SsKd$. Its elements are terminating sequences
\begin{align}
 f = (f^0,f^1,...,f^n,0,...)\,,
\end{align}
with $f^0\in\Cl^K,$ $f^n\in\SsKdn$. Addition, scalar multiplication and Fourier transformation is defined component wise, and we endow $\uSs$ with its usual topology \cite{BogolubovLogunovTodorov:1975}.

The Poincar\'e action \eqref{PGonBUA} can be extended to $\uSs$ by taking tensor products and direct sums, i.e. with $D=D(A^{-1})$, $\La=\La(A)$, we define
\begin{align*}
 f^n_{(y,A)}(x_1,...,x_n)_\sbk 
=
\sum_{l_1,...,l_n=1}^K D_{l_1k_1}\cdots D_{l_nk_n}
f^n(\La^{-1}(x_1-y),...,\La^{-1}(x_n-y))_\sbl
\,.
\end{align*}
The notations $f_{(y)}:=f_{(y,1)}$ and $\Psi(f):=\sum_n \Psi^n(f^n)$ for $f=(f^0,f^1,...,f^n,0,...,)\in\uSs$ will be used throughout.

Finally, we introduce two antilinear involutions $f\mapsto f^*$ and $f\mapsto f^J$ on $\uSs$,
\begin{align}
 (f^*)^n(x_1,...,x_n)_\sbk
&:=
\overline{f^n(x_n,...,x_1)_{\overline{\sbk}}}
\,,\qquad
\overline{\bk}=(\,\overline{k_n},...,\overline{k_1}\,)
\label{tomita-involution}
\,,\\
(f^J)^n(x_1,...,x_n)_\sbk
&:=
i^{N(\sbk)}\,\overline{f^n(-x_1,...,-x_n)_{\kbar_1...\kbar_n}}\label{tcp-involution}
\,,
\end{align}
related to the adjoint and TCP-transformed fields, respectively. Here $N(\bk):=\sum_{j=1}^n N(k_j)$ takes into account a possible spinorial character of the fields, with $N(k_j)\in\Zl$ and $N(\kbar_j)=N(k_j)$.

The vacuum expectation values of the fields, i.e. the $n$-point functions, are denoted by
\begin{align}
 \om_n(f^n) := \langle\Om,\,\Psi^n(f^n)\rangle\,,\qquad f^n\in\SsKdn\,,
\end{align}
and we also write $\om(f):=\sum_n\om_n(f^n)$. With these conventions, we have
\begin{align}
 \phi(f)\Psi(g)	&= \Psi(f\ot g)\,,\qquad f\in\Ss_K, g\in\uSs
\label{def:phi}
\,,\\
 \langle\Psi(g),\,\Psi(h)\rangle &= \om(g^*\ot h)\,,\qquad g,h\in\uSs\,.
\end{align}

\subsection{Noncommutative Minkowski space}

Having made precise our assumptions on the undeformed quantum field theory, let us describe the representation of the algebraic structure \eqref{NCMS} defining noncommutative Minkowski space which we will employ, following closely the original formulation in \cite{DoplicherFredenhagenRoberts:1995}.

The quantum conditions \eqref{NCMS} imply that the commutators $Q_{\mu\nu}=-i[X_\mu,X_\nu]$ commute with each other, and their joint spectrum is contained in the set
\begin{align}\label{Sigma}
\Sigma
:=
\Sigma_{\kae\kam}
=
\{\te\in\Rl^{4\times 4}_-\,:\, 
\te_{\mu\nu}\te^{\mu\nu}=2(\kae^2-\kam^2)\,,\;
\eps_{\mu\nu\alpha\beta}\te^{\mu\nu}\te^{\alpha\beta}= -8\,\kae\kam\}\,.
\end{align}
The two parameters $\kae, \kam$ entering into the construction will be taken as arbitrary but fixed real numbers in the following, and possible dependencies on these parameters will only be indicated when their values are of importance. The same convention applies to the reference matrix
\begin{align}\label{te1}
 \te_1
:=
\te_1(\kae,\kam)
=
\left(
\begin{array}{cccccc}
 0&\kae&0&0\\
 -\kae&0&0&0\\
0&0&0&\kam\\
0&0&-\kam&0
\end{array}
\right)
\in
\Sigma_{\kae\kam}
\,.
\end{align}
The set $\Sigma_{\kae\kam}$ is a homogeneous space for the proper orthochronous Lorentz group $\LGpo$ with respect to the action $\te\mapsto\La\te\La^T$, and to each $\te\in\Sigma_{\kae\kam}$ we associate a Lorentz transformation $\La_\te$ such that $\La_\te\te_1 \La_\te^T=\te$. 

For the formulation of a representation space for the commutation relations \eqref{NCMS}, we view $\Sigma$ as a submanifold of $\Rl^{16}$, and equip it with the corresponding differential structure and measure $d\sigma(\te)$. We then consider the Hilbert space
\begin{align}
 \VV := L^2(\Rl^2\times\Sigma, d^2s\times d\sigma(\te))
\end{align}
and its dense subspace $\VV^\infty := C_0^\infty(\Rl^2\times\Sigma)$.

Let $\xh_0,...,\xh_3$ denote the Schr\"odinger position and momentum operators acting on $L^2(\Rl^2, d^2s)$, i.e. $\hat{x}_0=s_1$, $\hat{x}_2=s_2$, $\hat{x}_1=-i\kae\,\partial_{s_1}$, $\hat{x}_3=-i\kam\,\partial_{s_2}$. Then $[\xh_\mu, \xh_\nu]=i(\te_1)_{\mu\nu}$, and the noncommutative coordinates $X_\mu$ are defined as, $v\in\VV^\infty$,
\begin{align}
 (X_\mu v)(s,\te) &:= ((\La_\te \hat{x})_\mu v)(s,\te)\,.
\end{align}
The commutators $Q_{\mu\nu}:=-i[X_\mu, X_\nu]$ satisfy $(Q_{\mu\nu}v)(s,\te)=\te_{\mu\nu}\cdot v(s,\te)$, and the joint spectrum of the $Q_{\mu\nu}$ is $\Sigma$. The $C^*$-algebra generated by the $X_\mu$ is denoted $\E$ and taken as the model of noncommutative Minkowski space \cite{DoplicherFredenhagenRoberts:1995}.

\subsection{Vacuum representations of the Weyl-Wigner deformed field algebra}

After these prerequisites, we turn to the formulation of the deformed quantum field theory. The basic idea for transporting the fields $\phi_k$ to noncommutative Minkowski space is to use a generalized Weyl-Wigner correspondence \cite{Szabo:2003}, i.e. to define ``$\phi_k(X)$'' with the help of the Fourier transform $\phiti_k$ of $\phi_k$, but making use of the exponentials $\exp(ip\cdot X)$ involving the noncommuting coordinates $X_\mu$. Due to the operator nature of the fields $\phi_k$, this correspondence is usually taken in the tensor product form $X\mapsto\int dp\, e^{ip\cdot X}\ot\phiti_k(p)$. Values of the field at different points are then defined with the help of the translations, which are implemented on the noncommutative Minkowski space by shifting the $X_\mu$ with multiples of the identity, $X_\mu\mapsto X_\mu+x_\mu\cdot 1$. These ideas can be summarized in the following formal definition \cite{DoplicherFredenhagenRoberts:1995, BahnsDoplicherFredenhagenPiacitelli:2005},
\begin{align}\label{heuristics}
 \phi^\ot_k(x)	&:= (2\pi)^{-2}\int d^4p\,\left(e^{ip\cdot X}\ot e^{ip\cdot x}\tilde{\phi}_k(p)\right)
\,,\qquad x\in\Rl^4
\,.
\end{align}
In the context of deformed free fields, the $\phiot_k(x)$ are usually considered as maps from states on $\E$ to field operators on Fock space \cite{BahnsDoplicherFredenhagenPiacitelli:2005}. We take here a slightly different point of view and want to study certain states and representations of the polynomial algebra generated by the fields $\phiot_k$. To this end, it is necessary to give rigorous meaning to the expression \eqref{heuristics} as a linear operator on some domain in $\VV\ot\Hil\cong L^2(\Rl^2\times\Sigma\to\Hil)$.

This can be done as follows. We consider the enlarged Borchers-Uhlmann algebra
\begin{align}
 \uSs^\ot:=C_0^\infty(\Rl^2\times\Sigma)\ot\uSs\,,
\end{align}
and denote its elements by bold face letters, $\bof=(\bof^0,\bof^1,...,\bof^n,0,...)$, and their dependence on $s, \te$ by subscripts, i.e. $\bof^n_{s,\te}\in \SsKdn$. These test functions are mapped to vectors in $L^2(\Rl^2\times\Sigma\to\Hil)$ via
\begin{align}
\bof\longmapsto \Psi^\ot(\bof)\,,\qquad\Psi^\ot(\bof)(s,\te) := \sum_n\Psi^n(\bof^n_{s,\te})\,.
\end{align}
The space spanned by the $\Psi^\ot(\bof)$ will be denoted $\DD^\ot$. To describe the fields $\phiot$, we define an action of $\SsKd$ on $\uSs^\ot$ by
\begin{align}
\widetilde{(f\times \bog)^n_{s,\te}}(p_1,...,p_n)_{k\sbl}
&:=
\fti(p_1)_k\cdot\widetilde{(e^{ip_1\cdot X}\bog)^{n-1}_{s,\te}}(p_2,...,p_n)_\sbl
\,,\quad f\in\SsKd, \;\bog\in\uSs^\ot\,.
\end{align}
Here the coordinates $X_\mu$ act in their previously defined Schr\"odinger representation on the $(s,\te)$-variables of $\bog$. Note that in view of the smooth and compactly supported $(s,\te)$-dependence of elements on $\uSs^\ot$, we have $f\times\bog\in\uSs^\ot$.

With this action, the Weyl-Wigner deformed quantum fields $\phiot$ take the form
\begin{align}
 \phiot(f)\Psi^\ot(\bog)
&=
\Psi^\ot(f\times \bog)
\,.
\end{align}
This formula can be regarded as the precise definition of the formal expression \eqref{heuristics}. Some of the relevant properties of the fields $\phiot(f)$ are summarized in the following proposition.

\begin{proposition}\label{prop1}
 The fields $\phiot(f)$ have the following properties:
\begin{enumerate}
 \item Each $\phiot(f)$, $f\in\SsKd$, is a well-defined linear operator on the dense domain $\DD^\ot$, and leaves $\DD^\ot$ invariant.
\item $\phiot(f)^*\supset\phiot(f^*)$.
\item The expectation value of products of fields in a vector state of the form $v\ot\Om$, $v\in\VV^\infty$, is
\begin{align}
\langle v\ot\Om,\; &\phiot(f_1)\cdots\phiot(f_n) \,v\ot\Om\rangle
\nonumber
\\
&=
\sum_\sbk \int ds\,d\sigma(\te) \int d^{4n}p\;
\omti_n(-p)_\sbk\,
\prod_{j=1}^n \fti_j(p_j)_{k_j}
\prod_{l<r}^n e^{-\frac{i}{2}p_l\te p_r}|v(s,\te)|^2
\,.\label{vecstate}
\end{align}
\end{enumerate}
\end{proposition}
\Proof
 a) To prove that $\phiot(f)$ is well defined, let $\bog\in\uSs^\ot$ with $\Psi^\ot(\bog)=0$, i.e.
\begin{align}\label{ginL}
\om(\bog_{s,\te}^*\ot\bog_{s,\te})=0\,,\qquad s\in\Rl^2,\;\, \te\in\Sigma\,.
\end{align}
Using the explicit Schr\"odinger representation of the $X_\mu$, one can easily show that given $f\in\uSs$, there exists $\boh\in\uSs^\ot$ (depending on $f, \bog$) such that
\begin{align*}
&\int ds\,d\sigma(\te)\,
\om((f\times\bog)_{s,\te}^*\ot(f\times\bog)_{s,\te})
=\int ds\, d\sigma(\te)
\,\om(\boh_{s,\te}\ot\bog_{s,\te})
\,.
\end{align*}
In view of \eqref{ginL} and the Cauchy-Schwarz inequality satisfied by $\om$,
\begin{align*}
 \left|
\int ds\, d\sigma(\te)\,
\om((f\times\bog)_{s,\te}^*\ot(f\times\bog)_{s,\te})
\right|
\leq
\int ds\, d\sigma(\te)
\,\om(\boh_{s,\te}^*\ot\boh_{s,\te})^{1/2}\om(\bog_{s,\te}^*\ot\bog_{s,\te})^{1/2}
=0\,.
\end{align*}
Since $\om$ is positive, this implies $\om((f\times\bog)_{s,\te}^*\ot(f\times\bog)_{s,\te})=0$ for all $s,\te$. Hence $\Psi^\ot(\bog)=0$ implies $\Psi^\ot(f\times\bog)=0$. Taking also into account that $f\times\bog\in\uSs^\ot$ for $\bog\in\uSs^\ot$, $f\in\SsKd$, it follows that $\phiot(f)$ is a well defined linear operator on the domain $\DD^\ot$, with $\phiot(f)\DD^\ot\subset\DD^\ot$.

b) With arbitrary $\bog,\boh\in\uSs^\ot$, $n,m\in\Nl_0$, $f\in\SsKd$, we compute
\begin{align*}
& \langle\Psi^\ot(\boh^n),\,\phiot(f)\Psi^\ot(\bog^m)\rangle
=
\int ds\,d\sigma(\te)\,
\om({\boh^n_{s,\te}}^*\ot(f\times\bog^m)_{s,\te})
\\
&=
\sum_{\sbk,l,\sbr}\int ds\,d\sigma(\te)
\int dq\,dq'\,dp\,
\omti_{n+m+1}(-q,-p,-q')_{\sbk l \sbr}
\overline{\tilde{\boh}^n_{s,\te}(-q_n,...,-q_1)_{\overline{\sbk}}}
\,\fti(p)_l(e^{ip\cdot X}\tilde{\bog}^m)_{s,\te}(q')_\sbr
\\
&=
\int ds\,d\sigma(\te)\,
\om((f^*\times\boh^n)_{s,\te}^*\ot \bog^m_{s,\te})
\\
&= \langle\Psi^\ot(f^*\times\boh^n),\,\Psi^\ot(\bog^m)\rangle
=
 \langle\phiot(f^*)\Psi^\ot(\boh^n),\,\Psi^\ot(\bog^m)\rangle\,.
\end{align*}
This implies $\phiot(f)^*\supset\phiot(f^*)$.

c) The vacuum vector $\Om\in\Hil$ is given by the constant function $1\in\uSs$ in the Borchers algebra, $1^n(x)=\delta_{n,0}$. By definition of $\phiot$, we therefore have
\begin{align}\label{phiotproduct}
 \phiot(f_1)\cdots\phiot(f_n)(v\ot\Om)=\Psi^\ot(\bof^n)\,,
\end{align}
with 
\begin{align}
\tilde{\bof}^n_{s,\te}(p_1,...p_n)_\sbk
&=
(f_1\times(f_2\times ... f_n\times(v\ot1)...))_{s,\te}(p_1,...,p_n)_\sbk
\nonumber
\\
&=
\fti_1(p_1)_{k_1}(e^{ip_1\cdot X}(f_2\times ...(f_n\times(v\ot1))...)_{s,\te}(p_2,...,p_n)
\nonumber
\\
&=
\prod_{j=1}^n\fti_j(p_j)_{k_j}
\cdot 
(e^{ip_1\cdot X}\cdots e^{ip_n\cdot X}v)(s,\te)\,.\label{bof-1}
\end{align}
Since the commutators $Q_{\mu\nu}=-i[X_\mu,X_\nu]$ act as $(Q_{\mu\nu}v)(s,\te)=\te_{\mu\nu}\cdot v(s,\te)$, it follows from the Baker-Campbell-Hausdorff formula that
\begin{align}
(e^{ip_1\cdot X}\cdots e^{ip_n\cdot X}v)(s,\te)
=
\prod_{1\leq l<r\leq n}e^{-\frac{i}{2}p_l\te p_r}
\cdot (e^{i\sum_{j=1}^np_j\cdot X}v)(s,\te)\,.
\end{align}
Putting these identities together, we arrive at the expectation values
\begin{align}
 \langle v&\ot\Om,\; \phiot(f_1)\cdots\phiot(f_n) \,v\ot\Om\rangle
=
\int ds\,d\sigma(\te)\,\langle v(s,\te)\cdot\Om\,,\Psi(\bof^n_{s,\te})\rangle
\label{exp1}
\\
&=
\sum_\sbk \int ds\,d\sigma(\te) \int d^{4n}p\;
\omti_n(-p)_\sbk\,
\prod_{j=1}^n \fti_j(p_j)_{k_j}
\prod_{l<r}^n e^{-\frac{i}{2}p_l\te p_r}\overline{v(s,\te)}(e^{i\sum_{j=1}^np_j\cdot X}v)(s,\te)
\,.\nonumber
\end{align}
In view of the translation invariance of the undeformed fields $\phi_k$, the $n$-point function $\omti_n(p)$ is non-vanishing only for zero total momentum $\sum_{j=1}^np_j$. We may therefore drop the factor $e^{i\sum_{j=1}^np_j\cdot X}$ in \eqref{exp1} and arrive at \eqref{vecstate}.
{\hfill  $\square$ \\[2mm]}

Regarding all polynomials in the fields $\phiot(f)$ and $\phiot(f)^*$ as being defined on $\DD^\ot$, we obtain a ${}^*$-algebra $\Pol^\ot$ of operators on $\VV\ot\Hil$, the algebra of Weyl-Wigner deformed quantum fields. In contrast to the commutative situation, $\Pol^\ot$ acts reducibly\footnote{We acknowledge a helpful discussion with E.~Seiler on this point.} on the domain $\DD^\ot$. In fact, $\Pol^\ot(v\ot\Om)$ is a nontrivial $\Pol^\ot$-stable subspace of $\VV\ot\Hil$, since $\Pol^\ot(u\ot\Om)\subset(\Pol^\ot(v\ot\Om))^\perp$ if $\int ds\,\overline{u(s,\te)}v(s,\te)=0$ for all $\te\in\Sigma$.

In the following, we will change to a different representation of $\Pol^\ot$, given via the GNS construction \cite{Schmudgen:1990} by a vacuum state on this algebra. Guided by the picture of modelling a situation in which the degrees of freedom of the noncommutative background are not coupled dynamically to the fields, we consider product states of the form $\nu\ot\om$ on $\Pol^\ot$, where $\om=\langle\Om,\,.\,\Om\rangle_\Hil$ is the vacuum state of the undeformed field algebra, and $\nu$ a (sufficiently regular) state on $\E$. To make contact with QFT on Moyal space, where the commutators of coordinates are taken to be multiples of the identity, $[X_\mu,X_\nu]=i\te_{\mu\nu}\cdot 1$, we will more specifically consider states $\om^\te:=\nu^\te\ot\om$, where $\nu^\te$ on $\E$ corresponds to a fixed spectral value\footnote{For a very different choice of state, see \cite{DoplicherFredenhagenRoberts:1995}.} $\te\in\Sigma$.

These states therefore have the form \eqref{vecstate} with $\int ds |v(s,\te')|^2$ replaced by $\delta(\te'-\te)$, i.e. 
\begin{align}\label{omte}
\om^\te(\phiot(f_1)\cdots\phiot(f_n))
=
\sum_\sbk \int d^{4n}p\;
\omti_n(-p)_\sbk\,
\prod_{j=1}^n \fti_j(p_j)_{k_j}
\prod_{l<r}^n e^{-\frac{i}{2}p_l\te p_r}\,.
\end{align}

For the description of the GNS representation of $\Pol^\ot$ with respect to $\om^\te$, we introduce the Moyal tensor product, a generalization of the star product to non-coinciding points \cite{Szabo:2003}.

\begin{definition}\label{def:mtp}{\bf (Moyal tensor product)}\\
Let $\te\in\Rl_-^{4\times 4}$. For $f^n\in\SsKdn$, $g^m \in\SsKdm$, the Moyal tensor product $f^n\otte g^m$ is defined via Fourier transformation as
\begin{align}\label{ttp}
\!\! (\widetilde{f^n\otte g^m})(p_1,...,p_n;q_1,...,q_m)_{\sbk\sbt}
	:=
 \prod_{l=1}^n\prod_{r=1}^m\,e^{-\frac{i}{2}p_l\te q_r}\cdot
 \fti^n(p_1,...,p_n)_\sbk\,\gti^m(q_1,...,q_m)_\sbt
\,.
\end{align}
By bilinearity of $\otte$, this definition is extended to $f=(f^0,f^1,...),\, g=(g^0,g^1,...)\in\uSs$.
\end{definition}

\begin{theorem}\label{thm-GNS}{\bf (Vacuum representations of $\Pol^\ot$ at fixed $\te\in\Sigma$)}\\
The GNS data of the Weyl-Wigner deformed field algebra $\Pol^\ot$ with respect to the state $\om^\te$ \eqref{omte} are up to unitary equivalence given by the Hilbert space $\Hil$ of the undeformed theory, the domain of definition $\DD$ \eqref{def:D}, and the vacuum vector $\Om$ as implementing vector, with the fields represented as, $f\in\SsKd$, $g\in\uSs$,
\begin{align}\label{solo}
 \pi^\te(\phiot(f))\Psi(g)=\Psi(f\otte g)\,.
\end{align}
\end{theorem}
\Proof
For comparison with the claimed formula \eqref{solo}, we define new fields $\phite(f)$, $f\in\Ss_K$, on $\DD\subset\Hil$ by $\phite(f)\Psi(g):=\Psi(f\otte g)$, $g\in\uSs$. In close analogy to the proof of Proposition \ref{prop1}, one can show that these are well-defined linear operators on $\DD$, which satisfy
\begin{align}\label{phitestar}
\phite(f)^*\supset\phite(f^*)\,.
\end{align}
For an explicit proof of these properties, see also Proposition \ref{prop-reg} below.

Furthermore, the vacuum $\Om$ is a cyclic vector for the polynomial algebra $\Pol^\te$ generated by all fields $\phite(f)$, $\in\Ss_K$, i.e. $\overline{\Pol^\te\Om}=\Hil$. To prove this claim, let $\Phi\perp\overline{\Pol^\te\Om}$ and note that an $n$-fold Moyal tensor product has the form $f_1,...,f_n\in\Ss_K$, 
\begin{align}
 \widetilde{(f_1\otte ... \otte f_n)}(p_1,...,p_n)_\sbk
&=
\prod_{j=1}^n \fti_j(p_j)_{k_j}\cdot \prod_{1\leq l<r\leq n}e^{-\frac{i}{2}p_l\te p_r}
\,.
\end{align}
Thus, for arbitrary $f_1,...,f_n\in\SsKd$,
\begin{align}
 0&=\langle \Phi,\,\phite(f_1)\cdots\phite(f_n)\Om\rangle 
=
\langle \Phi,\,\tilde{\Psi}^n(\tilde{F}^\te_n\cdot(\fti_1\ot...\ot \fti_n)\rangle\,,
\end{align}
where $\tilde{F}^\te_n(p_1,...,p_n)=\prod_{l<r}^n\exp(-\frac{i}{2}p_l\te p_r)$. By linearity and continuity, this equation also holds if $f_1\ot ... \ot f_n$ is replaced with any test function $g^n\in\SsKdn$. Choosing the Fourier transform of $g^n$ to have the form $(\tilde{F}^\te_n)^{-1}\cdot \hti^n$, with arbitrary $\hti^n\in\SsKdn$, we find $\langle\Phi,\,\Psi^n(h^n)\rangle=0$. In view of the cyclicity of the vacuum for the polynomial algebra of the undeformed fields, this implies $\Phi=0$ and hence $\overline{\Pol^\te\Om}=\Hil$.

Now let $(\DD^\te\subset\Hil^\te, \Om^\te, \pi^\te)$ denote the GNS data of the pair $(\Pol^\ot, \om^\te)$. We have to show that these data are unitarily equivalent to the ones given in the theorem. 
Since $\Om^\te$ is the GNS vector for $(\Pol^\ot,\om^\te)$, comparison with \eqref{omte} yields
\begin{align*}
 \langle\Om,\,\phite(f_1)\cdots\phite(f_n)\Om\rangle
=
\om^\te(\phiot(f_1)\cdots\phiot(f_n))
=
\langle\Om^\te,\,\pi^\te(\phi^\ot(f_1))\cdots \pi^\te(\phi^\ot(f_n))\Om^\te\rangle
\,.
\end{align*}
As a consequence of this identity, the coinciding form of the ${}^*$-structures (cf. Proposition \ref{prop1} b) and \eqref{phitestar}) for $\Pol^\ot$ and $\Pol^\te$, and the fact that $\Om^\te$ resp. $\Om$ is cyclic for $\pi^\te(\Pol^\ot)$ resp. $\Pol^\te$, it follows that the map $V:\Hil^\te\to\Hil$,
\begin{align}
 V\, \pi^\te(\phiot(f_1))\cdots\pi^\te(\phiot(f_n))\Om^\te := \phite(f_1)\cdots  \phite(f_n)\Om\,,
\end{align}
is well defined and maps $\pi^\te(\Pol^\ot)\Om^\te$ isometrically onto $\Pol^\te\Om$. Hence it extends to a unitary mapping $\Hil^\te$ onto $\overline{\Pol^\te\Om}=\Hil$.

By construction, $V\Om^\te=\Om$ and $V \pi^\te(\phi^\ot(f))V^* = \phite(f)$, so the proof of the theorem is finished.
{\hfill  $\square$ \\[2mm]}

\noindent{\it Remark:} Theorem \ref{thm-GNS} relies via Proposition \ref{prop1} c) on the translation invariance of the vacuum, but not on more specific properties of this state, like positivity of the energy or Lorentz invariance. It therefore also applies to more general translationally invariant states, such as thermal equilibrium states. 
\\
\\
Given the importance the $\te$-deformed fields introduced in the proof of Theorem \ref{thm-GNS} have in the subsequent sections, we write down their formal definition explicitly.

\begin{definition}\label{def:te-fields}{\bf ($\te$-deformed fields)}\\
 The $\te$-deformed quantum fields $\phite(f)$ are defined as linear operators on $\DD\subset\Hil$ by
\begin{align}\label{def:phite}
 \phite(f)\Psi(g) := \Psi(f\otte g)\,,\qquad f\in\SsKd,\,g\in\uSs\,.
\end{align}
The ${}^*$-algebra generated by these fields is denoted $\Pol^\te$.
\end{definition}
The fields $\phite$ have recently been found by Soloviev, who proposes them as a possible generalization of Wightman quantum fields to noncommutative Minkowski space \cite{Soloviev:2008}. Here we see how this construction is related to the choice of a particular vacuum state on the algebra of the Weyl-Wigner deformed fields $\phiot$.

For the case of a free field $\phi$, the $\te$-deformed fields $\phite$ have been constructed in our previous work \cite{GrosseLechner:2007}. Forgetting about the original motivation to understand representations of the Weyl-Wigner deformed field algebra and noncommutative Minkowski space, one can also view $\phi\to\phi^\te$ as a deformation of quantum field theories on the usual commutative Minkowski space. This point of view has been taken by Buchholz and Summers, who recently formulated a general algebraic version of this deformation, the warped convolution deformation \cite{BuchholzSummers:2008}. In the concrete Wightman setting discussed here, this deformation coincides with the replacement $\phi\to\phite$.
\\\\
By comparison of \eqref{def:phi} and Definition \ref{def:te-fields}, it becomes apparent that the deformation $\phi\to\phite$ amounts to introducing a new tensor product on the Borchers-Uhlmann algebra $\uSs$, replacing the usual tensor product $(f^n\ot g^m)(x,y)=f^n(x)g^m(y)$. Therefore, the properties of the $\te$-deformed fields can also be derived from properties of the Moyal tensor product, instead of following the more algebraic reasoning of \cite{BuchholzSummers:2008}. As we shall see later on, this has the advantage that more general deformations with interesting properties can be defined. In the following section, we extract the relevant properties of the tensor product $\otte$.

\section{The Moyal tensor product}\label{Sec:MTP}

The Moyal tensor product $\otte$ was introduced in Definition \ref{def:mtp} in momentum space, i.e. via Fourier transformation. Going over to position space, one checks by straightforward calculation that it can also be written as, $f^n\in\SsKdn$, $g^m\in\SsKdm$,
\begin{align}\label{ttp-x}
(f^n\otte g^m)(x,y)_{\sbk\sbt} 
&=
\pi^{-4}\int d^4\xi\int d^4q\,f^n_{(\xi)}(x)_\sbk\,g^m_{(\te q)}(y)_\sbt\,e^{-2i\xi\cdot q}\\
&=
\pi^{-4}\int d^4\xi\int d^4q\,f^n_{(-\te\xi)}(x)_\sbk\,g^m_{(q)}(y)_\sbt\,e^{-2i\xi\cdot q}\,.
\label{ttp-x-2}
\end{align}
For $n=m$ and scalar functions ($K=1$), one recovers the usual Moyal star product (with noncommutativity $\frac{1}{2}\te$) by passing to the diagonal $x=y$,
\begin{align}
 (f^n\otte g^n)(x,x) &= (f^n\star_{\te/2} g^n)(x)\,,\qquad f^n,g^n\in\Ssdn\,.
\end{align}
In the literature, sometimes also $f^n(x)\star g^m(y)$ is written instead of $(f^n\otte g^m)$ or $(f^n\otte g^m)(x,y)$, but for the sake of clarity, we stick to the tensor product notation $\otte$.

\begin{lemma}\label{lemma-mtp}{\bf (Basic properties of the Moyal tensor product)}\\
 Let $\te\in\Rl_-^{4\times 4}$. The corresponding Moyal tensor product $\otte$ has the following properties:
\begin{enumerate}
\item $\otte:\uSs\times\uSs\to\uSs$ is a bilinear, associative, and continuous map.
\item Equipping the space $\Rl_-^{4\times 4}$ of antisymmetric $(4\times 4)$-matrices with the matrix (operator) norm, the map $\te\mapsto f\otte g$ is continuous in the topology of $\uSs$ for any fixed $f,g\in\uSs$.
\item Poincar\'e transformations act on Moyal tensor products according to, $(y,A)\in\PGc$,
\begin{align}
(f\otte g)_{(y,A)}
&=
f_{(y,A)} \ot_{\La(A)\te\La(A)^T}g_{(y,A)}\,.
\end{align}
\item For $f,g\in\uSs$,
\begin{align}
(f\otte g)^* &= g^*\otte f^*\,,\\
(f\otte g)^J &= f^J\ot_{-\te} g^J \label{ttp-tcp}
\,.
\end{align}
\end{enumerate}
\end{lemma}
\Proof
 a) The bilinearity of $\otte$ follows directly from the definition \eqref{ttp}, and since multiplication with the function $\prod_{l<r}^ne^{-ip_l\te p_r/2}$ is a continuous map from $\Ss(\Rl^{4(n+m)})$ to $\Ss(\Rl^{4(n+m)})$, we also have continuity. Associativity, i.e. $f\otte(g\otte h)=(f\otte g)\otte h$, can easily be checked in momentum space.

b) This follows from Def. \ref{def:mtp} by using the estimate
\begin{align*}
 |e^{-\frac{i}{2}p\te q}-e^{-\frac{i}{2}p\te' q}| \leq \frac{1}{2}|p|\cdot|q|\cdot \|\te-\te'\|
\,,\qquad
p,q\in\Rl^4,\quad\te,\te'\in\Rl^{4\times 4}_-\,,
\end{align*}
and similar bounds for the derivatives of this function.

c) Note that since $p\te q=(p,\eta\te\eta q)$, there holds for any $\La\in\LG$ the equality
\begin{align*}
 (\La^{-1}p)\te (\La^{-1}q) = (\La^{-1}p,\eta\te\eta\La^{-1}q) =(p,\eta\La\te\La^T\eta q) = p \left(\La \te\La^T\right)q\,.
\end{align*}
The Fourier transform of $f^n_{(y,A)}$ is $\widetilde{f^n_{(y,A)}}(p)= e^{-i y\cdot\sum_{a=1}^n p_a}\cdot {\fti^n}_{\;\;\,(0,A)}(p)$. Writing $\La:=\La(A)$, this implies
\begin{align*}
 \widetilde{(f^n\otte g^m)_{(y,\La)}}(p,q)_{\sbk\sbt}
&=
	{\fti^n}_{\;\;(0,A)}(p)_\sbk
	\,
	{\gti^m}_{\;\;(0,A)}(q)_\sbt
	\,e^{-i y\cdot(\sum_{a=1}^{n}p_a+\sum_{b=1}^{m}q_b)}
	\prod_{l,r=1}^{n,m}
	e^{-\frac{i}{2}(\La^{-1}p_l)\te (\La^{-1}q_r)}
\\
&=
	\widetilde{f^n_{(y,A)}}(p)_\sbk
	\,
	\widetilde{g^m_{(y,A)}}(q)_\sbt
	\,
	\prod_{l,r=1}^{n,m}
	e^{-\frac{i}{2}p_l(\La\te\La^T)q_r}
\\
&=
\widetilde{(f^n_{(y,A)}\ot_{\La\te\La^T}g^m_{(y,A)})}(p,q)_{\sbk\sbt}\,,
\end{align*}
proving c).

d) The involutions $f\mapsto f^J$ and $f\mapsto f^*$ (\ref{tcp-involution}, \ref{tomita-involution}) act in momentum space according to
\begin{align}
 \widetilde{(f^J)^n}(p_1,...,p_n)_\sbk
&=
i^{N(\sbk)}\,\overline{\fti^n(p_1,...,p_n)_{\kbar_1..\kbar_n}}\,,\\
 \widetilde{(f^*)^n}(p_1,...,p_n)_\sbk
&=
\overline{\fti^n(-p_n,...,-p_1)_{\overline{\sbk}}}\,.
\end{align}
Since conjugation changes the sign of $\te$ in the phase factors $\exp(-\frac{i}{2}p_l\te p_r)$ in \eqref{ttp}, this implies $(f\otte g)^J=f^J\ot_{-\te}g^J$. For $f\mapsto f^*$, one has to take into account that inverting the order of the momenta amounts to exchanging $\te$ with $-\te$, too.
{\hfill  $\square$ \\[2mm]}

\noindent {\em Remark:} Whereas the Moyal tensor product $\otte$ is an associative product, brackets can {\em not} be omitted in multiple products if different noncommutativities $\te$, $\te'\in\Rl^{4\times 4}_-$ are involved, i.e. in general
\begin{align}
 (f\otte g)\ot_{\te'}h \neq f\otte(g\ot_{\te'}h)\,,\qquad \te\neq\te'\,.
\end{align}
\\
In our discussion of locality properties in Section \ref{Sec:Locality}, we will also need statements about support properties of Moyal tensor products, and therefore prove a corresponding proposition here. For its formulation, let us define the total momentum support of a function $f^n\in\Ssdn$ as
\begin{align}
 \Uti_f:=\{\,\sum_{j=1}^n p_j\,:\,(p_1,...,p_n)\in\supp\fti^n\,\}
\subset
\Rl^4\,,
\end{align}
and write for sets $S^n\subset\Rl^{4n}$
\begin{align}
S^n + \Uti_f
:=
\{(y_1+q,...,y_n+q)\,:\,(y_1,...,y_n)\in S^n\,,\;q\in\Uti_f\}\,.
\end{align}

\begin{proposition}\label{lemma-supp}{\bf (Support properties of Moyal tensor products)}\\
\vspace*{-12mm}
\\
\begin{enumerate}
 \item Let $f^n\in\Ssdn$, $g^m\in\Ssdm$. Then
\begin{align}
 \supp (f^n\otte g^m)\subset
	 \left(
\supp f^n+\tfrac{1}{2}\te\,\Uti_g\right)
	\times
\left(\supp g^m- \tfrac{1}{2}\te\,\Uti_f
	\right)\,.
\end{align}
\item Let $f_1,f_2\in\Ssd$, $g^n\in\Ssdn$ and consider a tempered distribution $\We\in\Ss(\Rl^{4(n+2)})'$ whose Fourier transform has support in the $(n+2)$-fold product of some cone $V\subset\Rl^4$. Then
\begin{align}\label{restrict}
 \We(f_1\otte(f_2\ot_{-\te}g^n))
=
\We(\chi_\U\cdot(f_1\otte(f_2\ot_{-\te}g^n)))
\,,
\end{align}
where $\chi_\U$ denotes the characteristic function of the set
\begin{align}\label{supp2}
 \U := (\supp f_1 -\te\, V) \times (\supp f_2 +\te\, V)\times\Rl^{4n}\,.
\end{align}
\item For $f_1,f_2\in\Ss(\Rl^4)$, $g^n\in\Ss(\Rl^{4n})$, 
\begin{align}\label{exchange-rule}
 (f_1\otte (f_2\ot_{-\te}g^n))(x_2,x_1,y)
=
(f_2\ot_{-\te} (f_1\otte g^n))(x_1,x_2,y)\,.
\end{align}
\end{enumerate}
\end{proposition}
\Proof
a) Let $\chiti_f$ and $\chiti_g$ denote the characteristic functions of $\Uti_f$ and $\Uti_g$, respectively. Since $\fti^n(p)=\chiti_f(\sum_{j=1}^n p_j)\cdot\fti^n(p)$, we can represent $f^n$ in \eqref{ttp-x} as a convolution with $\chi_f$:
\begin{align*}
 (f^n\otte g^m)(x,y)
&=
\frac{1}{4\pi^6}
\int d\xi\int dq\int dz\,
	f^n_{(\xi+z)}(x)\chi_f(z)\,
	g^m_{(\te q)}(y)
	e^{-2i\xi\cdot q}
\\
&=
\pi^{-4}
\int d\xi\int dq\,
	f^n_{(\xi)}(x)\chiti_f(-2q)\,g^m_{(\te q)}(y)
	e^{-2i\xi\cdot q}
\,.
\end{align*}
The $q$-integration is here restricted to $-\frac{1}{2}\,\Uti_f$ because of the support properties of $\chiti_f$. Thus $(f^n\otte g^m)(x,y)=0$ if $y$ is not contained in the set $(\supp g^m -\frac{1}{2}\te\,\Uti_f)$.

Alternatively, we can use \eqref{ttp-x-2} and represent $g^m$ as a convolution with $\chi_g$,
\begin{align*}
 (f^n\otte g^m)(x,y)
&=
\pi^{-4}
\int d\xi\int dq\,
	f^n_{(-\te\xi)}(x)\,g^m_{(q)}(y)\,\chiti_g(-2\xi)
	e^{-2i\xi\cdot q}\,,
\end{align*}
implying $(f^n\otte g^m)(x,y)=0$ for $x\notin(\supp f^n +\tfrac{1}{2}\te\,\Uti_g)$. This proves the claim about the support of $f^n\otte g^m$.

b) Let $\Uti_1, \Uti_2, \Uti_g$ denote the total momentum supports of $f_1, f_2, g$. By twofold application of a), 
the support $\cal S$ of $f_1\otte(f_2\ot_{-\te}g)$ can be estimated according to
\begin{align}
{\cal S}
&\subset
(\supp f_1 +\tfrac{1}{2}\te (\Uti_2+\Uti_g))
	\times
(\supp (f_2\ot_{-\te}g) -\tfrac{1}{2}\te\,\Uti_1)
\nonumber
\\
&\subset
(\supp f_1 +\tfrac{1}{2}\te (\Uti_2+\Uti_g))
	\times
(\supp f_2-\tfrac{1}{2}\te(\Uti_1+\Uti_g))
	\times
\Rl^{4n}\,.\label{supp-S}
\end{align}
For the evaluation of this test function in the distribution $\We$, we have to take into account that in
\begin{align*}
\We(f_1\otte(f_2\ot_{-\te}g^n))
&=
\int dp_1 dp_2 dq\,
\tilde{\We}(-p_1,-p_2,-q)
\fti_1(p_1)\fti_2(p_2)\gti^n(q)
e^{-\frac{i}{2}p_1\te(p_2+\hat{q})}
e^{\frac{i}{2}p_2\te \hat{q}}
,
\end{align*}
with $\hat{q}:=\sum_{j=1}^nq_j$, all momentum integrations are restricted to the cone $-V$. Hence we can proceed as in the proof of a), but use $-V$ instead of the total momentum supports to determine the restriction on the position space integrals in $\We(f_1\otte(f_2\ot_{-\te}g^n))$. After the replacement $\Uti_1,\Uti_2,\Uti_g\to-V$, the set \eqref{supp-S} turns into $\U$ \eqref{supp2} since $V$ is a cone. Thus \eqref{restrict} follows.

c) From Definition \ref{def:mtp} one can easily deduce the exchange rule, $f_1, f_2\in\Ssd$, $g^n\in\Ssdn$,
 \begin{align}
 \widetilde{(f_1\otte (f_2\ot_{\te'}g^n))}(p_2,p_1,q)
=
 e^{\frac{i}{2}p_1(\te+\te')p_2}
\cdot
\widetilde{(f_2\ot_{\te'} (f_1\otte g^n))}(p_1,p_2,q)
\,.
\end{align}
For the special case $\te'=-\te$, this simplifies to, $x_1,x_2\in\Rl^4$, $y\in\Rl^{4n}$,
\begin{align}
 (f_1\otte (f_2\ot_{-\te}g^n))(x_2,x_1,y)
=
(f_2\ot_{-\te} (f_1\otte g^n))(x_1,x_2,y)\,,
\end{align}
which is the claimed identity \eqref{exchange-rule}.
{\hfill  $\square$ \\[2mm]}

\section{$\te$-deformed quantum fields}\label{Sec:te-fields}

This section is devoted to the analysis of the $\te$-deformed quantum fields (Def. \ref{def:te-fields}), using the properties of the Moyal tensor product established before. In Subsection \ref{sec:reglim}, we consider the domain and continuity properties of the field operators $\phite(f)$, which turn out to be stable under the deformation. We also show that the commutative limit $\te\to0$ is continuous in a strong sense. In Subsection \ref{sec:npt}, we then discuss the structure of deformed $n$-point functions and comment on the reconstruction theorem in the deformed setting.

The most significant changes introduced by the noncommutative background are related to the covariance and locality properties of the quantum fields. These questions are considered in Subsection \ref{sec:covloc}.

References to the literature are given in the respective parts of this chapter.

\subsection{Regularity and commutative limit}\label{sec:reglim}
\begin{proposition}{\bf (Wightman properties of the deformed field operators)}\label{prop-reg}\\
Consider the $\te$-deformed fields $\phite(f)$ \eqref{def:phite}, with some noncommutativity $\te\in\Rl_-^{4\times 4}$. Then
\begin{enumerate}
 \item The dense subspace $\DD$ is contained in the domain of each $\phite(f)$, $f\in\SsKd$.
\item For $\Psi,\Psi'\in\DD$, the map $\SsKd\ni f \longmapsto \langle\Psi,\,\phite(f)\Psi'\rangle$ is a tempered distribution.
\item For any $f\in\SsKd$,
\begin{align}\label{star-phite}
 \phite(f)^*\supset \phite(f^*)\,.
\end{align}
\item For each open set $\OO\subset\Rl^4$, the space
\begin{align}
 \DD_\te(\OO) := {\rm span}\{\phite(f_1)\cdots\phite(f_n)\Om\,:\,f_j\in\Ss(\OO)^{\oplus K}\}
\end{align}
is dense in $\Hil$ (Reeh-Schlieder property).
\end{enumerate}
\end{proposition}
\Proof
 a) follows directly from the definition \eqref{def:phite} of $\phite(f)$, and b) is a consequence of the facts that the $\Psi^n$ are vector-valued distributions and $f\mapsto f\otte g$, $g\in\uSs$, is continuous in the Schwartz topology (Lemma \ref{lemma-mtp} a)).

c) This property has already been established by Soloviev \cite{Soloviev:2008}, but we give here a proof for the sake of self-containedness. To begin with, note that
\begin{align}
\om(f^*\ot g)
=
\om(f^*\otte g)
\,,\qquad f,g\in\uSs\,,
\end{align}
since the factor $\exp(-\frac{i}{2}(\sum_{l=1}^np_l)\te(\sum_{r=1}^m q_r))$ appearing in $(f^*)^n\otte g^m$ drops out because the Wightman distribution $\omti_{n+m}(p,q)$ has support in $\{(p,q)\,:\,\sum_{l=1}^np_l+\sum_{r=1}^m q_r=0\}$ and $\te$ is antisymmetric.

With $f\in\SsKd$, $g,h\in\uSs$, we therefore get
\begin{align*}
\langle\Psi(g),\,\phite(f)\Psi(h)\rangle
=
\langle\Psi(g),\,\Psi(f\otte h)\rangle
=
\om(g^*\ot(f\otte h))
=
\om(g^*\otte f\otte h)
\,.
\end{align*}
Making use of Lemma \ref{lemma-mtp} d), we furthermore see
\begin{align*}
\om(g^*\otte f\otte h)
=
\om((f^*\otte g)^*\otte h)
=
\langle\Psi(f^*\otte g),\,\Psi(h)\rangle
=
\langle\phite(f^*)\Psi(g),\,\Psi(h)\rangle
.
\end{align*}
This proves \eqref{star-phite}.

For d), we first observe that $\DD_\te(\Rl^4)=\pi^\te(\Pol^\ot)\Om$ is dense in $\Hil$ because $\Om$ is a cyclic vector for the GNS representation $\pi^\te$. To establish the density of the restricted spaces $\DD_\te(\OO)$, with $\OO\subset\Rl^4$ some open set, note that $\phite(f)$ transforms covariantly under translations,
\begin{align}
 U(x,1)\phite(f)U(x,1)^{-1}
&=
\phite(f_{(x)})\,,\qquad f\in\SsKd\,.
\end{align}
(For more general transformation properties of the fields $\phite$, see also Lemma \ref{lemma-trans} below.) In view of the undeformed spectral properties of the translation group (positivity of the energy), we can now apply the usual Reeh-Schlieder argument \cite{StreaterWightman:1964} to conclude the density of $\DD_\te(\OO)\subset\Hil$ from that of $\DD_\te(\Rl^4)\subset\Hil$.
{\hfill  $\square$ \\[2mm]}

The undeformed fields $\phi_k$ are included in our considerations as the special case $\te=0$. Moreover, one can recover the original theory in the limit $\te\to0$ of vanishing noncommutativity. This continuity of the $\te$-deformation is proved next.

\begin{proposition}{\bf (The commutative limit)}
The $\te$-deformed field operators $\phite(f)$ converge strongly to the undeformed field operators $\phi(f)$ on $\DD$ as $\te\to 0$.
\end{proposition}
\Proof
Let $f\in\SsKd$ and $g\in\uSs$. Taking into account that the $\Psi$ are vector-valued tempered distributions, the continuity of $\te\mapsto f\otte g$ established in Lemma \ref{lemma-mtp} b) implies
\begin{align}
 \lim_{\te\to0}\phite(f)\Psi(g)
=
 \lim_{\te\to0}\Psi(f\otte g)
=
\Psi(f\ot g)
=
\phi(f)\Psi(g)\,.
\end{align}
Since any vector in $\DD$ is of the form $\Psi(g)$ for some $g\in\uSs$, this proves the claim.
{\hfill  $\square$ \\[2mm]}

\subsection{Deformed $n$-point functions and reconstruction}\label{sec:npt}

A Wightman quantum field theory can be completely characterized in terms of its $n$-point functions $\om_n(x_1,...,x_n)_\sbk=\langle\Om,\,\phi_{k_1}(x_1)\cdots \phi_{k_n}(x_n)\Om\rangle$ \cite{StreaterWightman:1964}. In the deformed setting we consider here, the (smeared) vacuum expectation values of products of fields are
\begin{align}
 \om^\te(f_1\ot...\ot f_n) 
&:=
\langle\Om\,,\phite(f_1)\cdots\phite(f_n)\,\Om\rangle 
= 
 \langle\Om\,,\Psi^n(f_1\otte ...\otte f_n)\rangle
\nonumber
\\
&=
\om(f_1\otte ...\otte f_n)\,,
\label{omte1}
\end{align}
In the following, we will use the notation $\om^\te(f)=\sum_n\om^\te_n(f^n)$, $f=(f^0,f^1,...,0,...)\in\uSs$, with the distributions $\om^\te_n$ defined by linear and continuous extension of \eqref{omte1} to $\SsKdn$, and write $\om_n^0:=\om_n$ to emphasize the undeformed $n$-point functions. 

 The distributional kernels of the deformed Wightman functions have in momentum space the universal $\te$-dependence
\begin{align}
\tilde{\om}_n^\te(p_1,...,p_n)_\sbk
=
\prod_{1\leq l<r\leq n}e^{-\frac{i}{2}p_l\te p_r}
\cdot
\tilde{\om}_n^0(p_1,...,p_n)_\sbk
\,.
\end{align}
Definition \ref{def:te-fields} implies that $\phite(f)\Om=\phi(f)\Om$ does not depend on $\te\in\Rl^{4\times 4}_-$. Hence the vacuum expectation value of a single field and the two-point function are always undeformed, $\om^\te_1(x_1)=\om^0_1(x_1)$, $\om^\te_2(x_1,x_2)=\om^0_2(x_1,x_2)$. The twisting factor $\prod_{l<r}e^{-\frac{i}{2}p_l\te p_r}$ introduces a non-trivial $\te$-dependence only in the higher $n$-point functions, $n\geq3$.
\\\\
As pointed out by Soloviev \cite{Soloviev:2008}, the inner products, $f,g\in\uSs$,
\begin{align}
 (f,g)_\te
&:=
\om^\te(f^*\ot g)
=
\om^\te(f^*\otte g)
\end{align}
are positive semi-definite, i.e. we can use them in the same way as in the reconstruction theorem of Wightman theory \cite{StreaterWightman:1964} to define different Hilbert space structures on the Borchers-Uhlmann algebra $\uSs$. Calling the completed spaces $\Hil_\te$, $\te\in\Rl^{4\times4}_-$, we consider the maps $u_\te:\uSs\to\uSs$, defined by
\begin{align}
\widetilde{(u_\te f)^n}(p)
&:=
\prod_{1\leq l<r\leq n}e^{-\frac{i}{2}p_l\te p_r}\cdot \widetilde{f^n}(p)
\,.
\end{align}
With the help of the translation invariance of the $\om_n^0$, one readily proves
\begin{align*}
(u_\te f, u_\te g)_0 
=
\om^0((u_\te f)^*\ot u_\te g)
=
\om^0(u_\te f^*\otte u_\te g)
=
\om^\te(f^*\ot g)
=
(f,g)_\te
\,,
\end{align*}
i.e. the $u_\te$ extend to unitaries $U_\te$ mapping $\Hil_\te$ onto $\Hil_0$, with ${U_\te}^{-1}=U_{-\te}$. To reconstruct the field operators, we consider the maps $\varphi_\te(f):\uSs\to\uSs$, $f\in\SsKd$,
\begin{align}
 \varphi_\te (f)g := f\otte g
\,,
\end{align}
which are intertwined by the $u_\te$, i.e. $\varphi_\te(f)=u_\te\,\varphi_0(f)u_{\te}^{-1}$. In the same way as in the proof of Proposition \ref{prop1} a), one can show that $\{g\in\uSs\;:\;(g,g)_0=0\}$ is a left ideal with respect to the Moyal tensor product. Hence the maps $\varphi_\te(f)$ give rise to linear operators on $\Hil_0=\Hil$ via the usual Wightman reconstruction procedure -- these are the $\te$-deformed fields $\phite(f)$ introduced in Definition \ref{def:te-fields}.

One can also consider $\varphi_0(f)$ as an operator on $\Hil_\te$, but the relation between $\varphi_\te(f)$ and $\varphi_0(f)$ implies that the latter point of view is unitarily equivalent to the former. We work here with $f\mapsto f\otte g$ on $\Hil_0$ in order to represent all fields $\phite$, $\te\in\Rl^{4\times 4}_-$, on the same Hilbert space.
\\
\\
For the case of a scalar neutral free field $\phi_o$, the deformed field operator $\phite_o$ can be described in terms of twisted creation/annihilation operators $a^\#_\te(p)$ \cite{GrosseLechner:2007}. But also for general Wightman fields $\phi$, one can specify the distributional kernels of the deformed fields $\phite(f)$ explicitly. With $f\in\SsKd$, $g^n\in\SsKdn$, we have
\begin{align*}
\phite(f)\Psi^n(g^n)
=
\sum_{k,\sbl}
\int d^4p\,
\fti(-p)_k
\int d^{4n}q\,
\gti^n(-q)_\sbl\,e^{-\frac{i}{2}p\te \sum_{j=1}^n q_j}
\,
\phiti_k(p)\,\Psiti^n(q)_\sbl
\,.
\end{align*}
Since the vectors $\Psiti^n(q)_\sbl$ are eigenvectors of the energy-momentum operators $P^\mu$, with eigenvalues $\sum_{j=1}^n q_j^\mu$, the kernels of the deformed fields can be written as
\begin{align}\label{phite-kernel}
 \phitite_k(p) = \phiti_k(p)\,e^{-\frac{i}{2}p\te P} =  e^{-\frac{i}{2}p\te P}\,\phiti_k(p)\,.
\end{align}
The second equality follows from the translation covariance of $\phi_k$ and the antisymmetry of $\te$. In position space, \eqref{phite-kernel} formally reads
\begin{align}
 \phite_k(x) = \exp\bigg(-\frac{1}{2}\frac{\partial}{\partial x^\mu}\,\te^{\mu\nu}\,P_\nu\bigg)\,\phi_k(x)\,,
\end{align}
a formula which has been used in the work of Balachandran et. al. for  deformed free fields \cite{BalachandranPinzulQureshiVaidya:2007}. Definition \ref{def:te-fields} can be understood as a way of giving rigorous meaning to this formal expression for general quantum fields.

\subsection{Covariance and TCP properties}\label{sec:covloc}

As we saw in the previous section, the deformed fields $\phite_k$ do not differ much from the undeformed Wightman fields $\phi_k$ as  far as domain and continuity properties are concerned. However, the noncommutative background is expected to lead to drastic changes in comparison to the commutative case when it comes to questions of covariance and localization. The vacuum state $\om^\te$ fixes a specific value $\te$ in the joint spectrum of the commutators $[X_\mu, X_\nu]=i\,Q_{\mu\nu}$ and thus breaks Lorentz invariance to a subgroup if $\te\neq 0$. This feature manifests itself here in a modified transformation rule for the deformed fields.

\begin{lemma}\label{lemma-trans}{\bf (Poincar\'e transformation properties of the deformed fields)}\\
Let $(a,A)\in\PGc$ and $f\in\Ssd$. Then, with $\La=\La(A)$,
\begin{align}\label{phi-trans}
 U(a,A)\phite(f)U(a,A)^{-1}\,\Psi
&=
\phi^{\La\te\La^T}(f_{(a,A)})\,\Psi\,,\qquad \Psi\in\DD\,.
\end{align}
\end{lemma}
\Proof
Let $g\in\uSs$. Then
\begin{align}\label{covlem1}
 U(a,A)\phite(f)U(a,A)^{-1}\Psi(g)
&=
U(a,A) \Psi(f\otte g_{(a,A)^{-1}})
\nonumber \\
&=
\Psi((f\otte g_{(a,A)^{-1}})_{(a,A)})
\nonumber \\
&=
\Psi(f_{(a,A)}\ot_{\La\te\La^T} g)
\nonumber \\
&=
\phi^{\La\te\La^T}(f_{(a,A)})\Psi(g)
\,,
\end{align}
where we used Lemma \ref{lemma-mtp} c) in the third equality.
{\hfill  $\square$ \\[2mm]}

From the point of view of an observer preparing a state of the form $\om^\te$, with some $\te\in\Sigma_{\kae\kam}$, rotated or boosted systems appear in different states $\om^{\te'}$, with the noncommutativity parameter $\te'$ varying over the orbit $\Sigma_{\kae\kam}$. Each $\te$-deformed field $\phite$ transforms covariantly only under those Lorentz transformations $\La$ which satisfy $\La\te\La^T=\te$. If both the parameters $\kae, \kam$ labelling the orbit \eqref{Sigma} are different from zero, this subgroup is ${\rm SO}(1,1)\times {\rm SO}(2)$, with the two factors corresponding to boosts in the $x_1$-direction and rotations in the $x_2$-$x_3$-plane in the case of the reference matrix $\te_1$ \eqref{te1}.

However, the model given by the {\em family} of fields $\{\phite(f)\,:\,\te\in\Sigma_{\kae\kam},\, f\in\SsKd\}$ is covariant under the full group $\PGc$, with the modified transformation law \eqref{phi-trans}. This field theory does not depend on a specific value of the noncommutativity parameter $\te$, but only on a chosen Lorentz orbit $\Sigma_{\kae\kam}\subset\Rl^{4\times 4}_-$, i.e. on the two parameters $\kae,\kam\in\Rl$.
\\
\\
In usual Wightman quantum field theory, it is well known that the representation $U$ of $\PGc$ can be extended by an antiunitary TCP operator $J$ implementing spacetime reflection and charge conjugation. The undeformed fields transform covariantly under this operator, i.e.
\begin{align}
 J\phi(f)J^{-1}
&=
\phi(f^J)
\,,\qquad
J\Psi(g)
=
\Psi(g^J)
\,,
\end{align}
where $g\to g^J$ denotes the involution \eqref{tcp-involution}.

In many models, even stronger covariance properties are realized, and all Poincar\'e transformations act as symmetries. In particular, time reflection $r_T(x_0,\bx):=(-x_0,\bx)$ and space reflection $r_P(x_0,\bx):=(x_0,-\bx)$ are then represented by (anti-) unitary operators $T:=U(0,r_T)$ and $P:=U(0,r_P)$ (not to be confused with the energy-momentum operators $P_0,...,P_3$) such that, $f\in\Ssd$,
\begin{align}
 T\phi_k(f)T^{-1}
&=
\alpha_T(k)\cdot
\phi_k(\overline{f_{(0,r_T)}})
\,,\qquad
P\phi_k(f)P^{-1}
=
\alpha_P(k)\cdot
\phi_k(f_{(0,r_P)})\,,
\end{align}
with phases $\alpha_T(k), \alpha_P(k)\in\{1,-1,i,-i\}$. In this case, there also exists a charge conjugation operator $C$, $C\phi_k(f)C^{-1}=\alpha_C(k)\cdot \phi_{\kbar}(f)$, with $\alpha_C(k)\alpha_P(k)\alpha_T(k)=i^{N(k)}$, and $J$ coincides with the product $TCP$ \cite{BogolubovLogunovTodorov:1975}.

In the $\te$-deformed framework considered here, the situation looks as follows.
\begin{proposition}\label{Prop:TCP}{\bf (TCP and reflection symmetries for deformed quantum fields)}\\
The TCP transformation $J$ acts on the deformed fields according to
\begin{align}
 J\phite(f)J^{-1} = \phi^{-\te}(f^J)\,.
\end{align}
If the transformations $P$, $C$ and $T$ are realized separately as symmetries of the undeformed theory, the deformed fields satisfy
\begin{align}
P\phite_k(f)P^{-1}
&=
\alpha_P(k)\cdot
\phi^{r_P\te r_P}_k(f_{(0,r_P)})
\,,\label{P-te}
\\
C\phite_k(f)C^{-1}
&=
\alpha_C(k)\cdot
\phite_{\kbar}(f)
\,,\label{C-te}
\\
T\phite_k(f)T^{-1}
&=
\alpha_T(k)\cdot
\phi^{-r_T\te r_T}_k(\overline{f_{(0,r_T)}})
\,.\label{T-te}
\end{align}
\end{proposition}
\Proof
 We first use Lemma \ref{lemma-mtp} c) to compute the action of $J$. With $f\in\SsKd$, $g\in\uSs$, there holds
\begin{align*}
 J\phite(f)J\Psi(g)
&=
J\Psi(f\otte g^J)
=
\Psi(f^J\ot_{-\te}g)
=
\phi^{-\te}(f^J)\Psi(g)\,.
\end{align*}

The proof of Lemma \ref{lemma-trans} can immediately be extended to cover also the parity transformation $P$, leading to \eqref{P-te}. For time reflection, one has to take into account that $T$ is antilinear: This conjugation flips $\te$ to $-\te$ and also leads to a conjugation of the testfunction in $T\phite_k(f)T^{-1}$ \eqref{T-te}. Finally, for charge conjugation we have $C\Psi^n(g^n)=\sum_\sbk\alpha_C(k_1)\cdots\alpha_C(k_n)\Psi^n(g^n_{\overline{\sbk}})$, which implies \eqref{C-te}.
{\hfill  $\square$ \\[2mm]}

Note that with $\te\in\Sigma_{\kae\kam}$, also $-\te$ lies on this orbit, i.e. the TCP transformed fields $J\phite(f)J^{-1}$ are also elements of the polynomial algebra generated by the fields $\phite(f)$, $\te\in\Sigma_{\kae\kam}$, $f\in\SsKd$. In the following, we will only use the (cover of) the proper Poincar\'e group as symmetry group, since the individual reflections $T$, $C$, and $P$ might already be broken on the level of the undeformed theory.
\\
\\
The TCP theorem in the context of $\te$-deformed theories has attracted some attention in the literature \cite{ChaichianNishijimaTureanu:2003, AlvarezGaumeVazquezMozo:2003,AkoforBalachandranJoJoseph:2007}. In \cite{AlvarezGaumeVazquezMozo:2003}, a different model-independent setup for Wightman theories on Moyal space was proposed, with the essential ingredient that the geometric symmetry group is ${\rm O}(1,1)\times {\rm SO}(2)$. This kind of symmetry is shared by the algebra of fields $\phite$ belonging to a fixed $\te$ in a Lorentz orbit $\Sigma_{0,\kam}$ with $\kae=0$ in our setting, i.e. on a Moyal space with ``commuting time''.  In \cite{AlvarezGaumeVazquezMozo:2003}, the authors consider this weakened symmetry together with a weakened locality assumption (cf. also the discussion after Theorem \ref{thm:wedge-locality} below), which can be used to derive the TCP theorem in that setting, see also \cite{ChaichianNishijimaTureanu:2003} for a somewhat similar approach.

In the special case of deformations of the free scalar neutral field, the TCP symmetry of Proposition \ref{Prop:TCP} has been established before \cite{AkoforBalachandranJoJoseph:2007,GrosseLechner:2007}.

\subsection{Localization in wedges}\label{Sec:Locality}

It is well known that quantum field theory on noncommutative spacetimes typically exhibits nonlocal features. In general, one has for $f,g\in\Ss(\Rl^4)$ with space-like separated supports, $\supp f\subset(\supp g)'$,
\begin{align}
 [\phite_k(f),\phite_l(g)]_\pm \neq 0\,,
\end{align}
in contrast to the undeformed situation at $\te=0$ \eqref{locality}. At small scales of the order of magnitude of the Planck length, such a violation of locality might be acceptable from a physics point of view, but at larger scales, nonlocality has to be regarded as an unphysical artifact of the chosen model.

We therefore want to investigate in the following to which degree locality is broken in our setting, and will find a weakened concept of localization which is still compatible with noncommutativity. As for the Lorentz transformation properties, our point of view is that of an observer preparing a vacuum state $\om^\te$ with sharp noncommutativity parameter $\te\in\Sigma_{\kae\kam}$. The question we consider is if it is possible to consistently assign localization regions $\OO\subset\Rl^4$ (presumably larger than a single point set $\{x\}$) to the field operators $\phite(x)$, such that the transformed fields $U(a,A)\phite(x)U(a,A)^{-1}$ commute with $\phite(x)$ whenever $\La(A)\OO+a$ lies spacelike to $\OO$.

In the context of a deformed free field $\phite_o$, it has been shown that although the point-like localization of $\phi_o$ is lost for $\te\neq0$, the fields $\phite_o$ are localized in certain infinitely extended, wedge-shaped regions of Minkowski space  \cite{GrosseLechner:2007}: For any $\te\in\Sigma$, there exists a wedge region $W(\te)\subset\Rl^4$ such that $\phite_o(x)$ is localized in $W(\te)+x$ in the above mentioned sense.

The same type of localization was also found in the generalized deformation studied by Buchholz and Summers \cite{BuchholzSummers:2008}. Here we show how the wedge-locality of the $\te$-deformed fields can be derived from properties of the Moyal tensor product, and first recall some facts about wedges.
\\
\\
As our reference region, we take the {\em standard wedge} $W_1$ in $x_1$-direction,
\begin{align}\label{W1}
 W_1 &:= \{x\in\Rl^4\,:\,x_1>|x_0|\}\,,
\end{align}
and the {\em set of all wedges} is defined to consist of all Lorentz transforms of $W_1$,
\begin{align}
 \W_0:=\LGpo W_1
=
\{\La W_1\,:\,\La\in\LGpo\}
\,.
\end{align}
It has been shown in \cite{GrosseLechner:2007} that the sets $\Sigma_{\kae\kam}$ and $\W_0$ are homomorphic as homogeneous spaces for the proper Lorentz group since the stabilizer group of $\te_1$ \eqref{te1} with respect to the action $\te\mapsto\La\te\La^T$ and the stabilizer group of $W_1$ \eqref{W1} with respect to the action $W\mapsto\La W$ coincide if $\kae\neq0$, $\kam\neq0$. This also holds if we consider $\Sigma$ and $\W_0$ as homogeneous spaces for the proper Lorentz group $\LGp$ and represent the spacetime reflection by $\te\mapsto-\te$ (cf. Prop. \ref{Prop:TCP}).

We can therefore associate a wedge $W(\te)\in\W_0$ with each $\te\in\Sigma$ in a covariant manner by putting
\begin{align}
 W(\La\te_1\La^T) := \eps(\kae)\,\La W_1\,,\qquad \La\in\LGpo\,;
\end{align}
this assignment is well defined in view of the above remarks. Here $\eps(\kae)$ denotes the sign of the parameter $\kae$ appearing in the definition of the orbit $\Sigma=\Sigma_{\kae\kam}$ \eqref{Sigma}, i.e. $\te_1$ is associated with $W_1$ if $\kae\geq0$, and $\te_1$ is associated with $-W_1$ if $\kae<0$.

With the convention \eqref{te-conv}, the reference noncommutativity $\te_1$ \eqref{te1} maps the positive lightcone into the wedge $-W_1$ if $\kae\geq0$, i.e. $\te_1\,V_+\subset -\overline{W(\te_1)}$. 

The causal complement of $W_1$ is $W_1'=-W_1$. Since spacetime reflection $j:x\mapsto-x$ is implemented by $\te\mapsto-\te$ on $\Sigma_{\kae\kam}$, we have
\begin{align}
 W(\te_1)'=-W(\te_1)=jW(\te_1)=-W(\te_1)\,.
\end{align}
By standard covariance arguments, these remarks imply the following facts. For a), see for example \cite{ThomasWichmann:1997}.
\begin{itemize}
\item[W1)] Let $W_1,W_2\in\W_0$. Then $W_1\subset W_2 \Longleftrightarrow W_1=W_2$.
\item[W2)] The causal complement of a wedge $W\in\W_0$ is $W'=-W$.
\item[W3)] $W(\te)=-W(\te') \Longleftrightarrow \te=-\te'$.
\item[W4)] $\te\, V_+ \subset -\overline{W(\te)}\,,\qquad\te\in\Sigma\,.$
\end{itemize}
Our theorem regarding the localization of the fields $\phite_k$ reads as follows:

\begin{theorem}\label{thm:wedge-locality}{\bf ($\te$-deformed quantum fields are wedge-local)}\\
If two undeformed fields  $\phi_k$, $\phi_l$ commute or anticommute at spacelike separation,
\begin{align}\label{loc2}
 [\phi_k(x),\,\phi_l(y)]_\pm =0\,,\qquad (x-y)^2<0\,,
\end{align}
then their $\te$-deformed counterparts satisfy the following wedge-local (anti-) commutation relations: For $\Psi\in\DD$, noncommutativity parameters $\te,\te'\in\Sigma_{\kae\kam}$, and test functions $f_1, f_2\in\Ssd$ with
\begin{align}\label{W-causal}
 \supp f_1 + W(\te) \subset \left( \supp f_2 + W(\te') \right)'\,,
\end{align}
there holds
\begin{align}
[\phite_k(f_1),\,\phi^{\te'}_l(f_2)]_\pm\Psi = 0\,.
\end{align}
Hence the fields $\phite_k(x)$ are localized in the wedge regions $W(\te)+x$.
\end{theorem}
\Proof
In view of the above remarks W1)-W3) on the structure of $\W_0$, the condition of spacelike separation \eqref{W-causal} implies $W(\te')=W(\te)'=-W(\te)$, and hence $\te'=-\te$. So we consider indices $k,l\in\{1,...,K\}$ such that \eqref{loc2} holds, and testfunctions $f_1,f_2\in\Ss(\Rl^4)$ with $\supp f_1 + W(\te) \subset \left( \supp f_2 - W(\te) \right)'$. But this inclusion only occurs for test functions $f_1,f_2$ such that there exists some translation $a\in\Rl^4$ with $\supp f_1+a\subset W(\te)$ and $\supp f_2+a\subset -W(\te)$. In view of the translation covariance of the theory, it therefore suffices to consider the case $\supp f_1\subset W(\te)$, $\supp f_2\subset -W(\te)$ for the proof of the theorem.

Choosing such $f_1,f_2$, an arbitrary $g^n\in\Ssdn$ and a multi-index $\bm$, we get
\begin{align*}
 [\phite_k(f_1), \phi^{-\te}_l(f_2)]_\pm \Psi^n_\sbm(g^n)
=
\Psi^{n+2}_{kl\sbm}(f_1\otte(f_2\ot_{-\te}g^n)) \pm \Psi^{n+2}_{lk\sbm}(f_2\ot_{-\te}(f_1\otte g^n))
.
\end{align*}
With the help of the exchange relation \eqref{exchange-rule},
\begin{align}
(f_2\ot_{-\te} (f_1\otte g^n))(x_2,x_1,y)
=
 (f_1\otte (f_2\ot_{-\te}g^n))(x_1,x_2,y)
\,,
\end{align}
we can express the (anti-) commutator of the deformed fields in terms of the (anti-) commutator of the undeformed fields as
\begin{align*}
 [\phite_k(f_1), \phi^{-\te}_l(f_2)]_\pm \Psi^n_\sbm(g^n)
&=
\int dx_1 dx_2 dy\, (f_1\otte (f_2\ot_{-\te}g^n))(x_1,x_2,y)\,[\phi_k(x_1), \phi_l(x_2)]_\pm \Psi^n_\sbm(y).
\end{align*}
For any vector $\Phi\in\Hil$, we therefore have
\begin{align}\label{comm-dist}
  \langle\Phi,\,[\phite_k(f_1), \phi^{-\te}_l(f_2)]_\pm \Psi^n_\sbm(g^n)\rangle
&=
\We(f_1\otte (f_2\ot_{-\te}g^n))\,,
\end{align}
where $\We$ denotes the distribution with kernel $\We(x_1,x_2,y)=\langle\Phi,\,[\phi_k(x_1), \phi_l(x_2)]_\pm \Psi^n_\sbm(y)\rangle$.

The Fourier transform of $\We$ has support in the $(n+2)$-fold product of the forward light cone $V_+$ as a consequence of the spectrum condition. We are thus in the position to apply Proposition \ref{lemma-supp} b), which yields
\begin{align}\label{comm-dist2}
  \langle\Phi,\,[\phite_k(f_1), \phi^{-\te}_l(f_2)]_\pm \Psi^n_\sbm(g^n)\rangle
&=
\We(\chi_\U\cdot(f_1\otte (f_2\ot_{-\te}g^n)))\,,
\end{align}
where $\chi_\U$ is the characteristic function of the set \eqref{supp2}
\begin{align*}
 \U &= (\supp f_1-\te V_+)\times(\supp f_2+\te V_+)\times \Rl^{4n}
\\
&\subset
(\supp f_1+\overline{W(\te)})\times(\supp f_2-\overline{W(\te)})\times\Rl^{4n}
\\
&\subset
 W(\te)\times W(\te)'\times \Rl^{4n}
\,.
\end{align*}
In the second line, we used the inclusion property W4), and in the third line the. support properties of $f_1$ and $f_2$.

From this form of $\U$, we see that for all $(x_1,x_2,y)\in\U$, $x_1$ lies spacelike to $x_2$. But the commutator distribution $\We$ vanishes for spacelike separated $x_1,x_2$ in view of the locality of the undeformed fields. So we arrive at
\begin{align}
 \langle\Phi,\,  [\phite_k(f_1), \phi^{-\te}_l(f_2)]_\pm \Psi^n_\sbm(g)\rangle=0\,,
\end{align}
and since $\Phi$, $g$, $n$ and $\bm$ were arbitrary, the statement of the theorem follows.
{\hfill  $\square$ \\[2mm]}

The localization properties of various approaches to noncommutative quantum field theories have been discussed in the literature before, and we would like to point out a difference between the approach taken by \'Alvarez-Gaum\'e and V\'azquez-Mozo and our formulation  \cite{AlvarezGaumeVazquezMozo:2003}. These authors consider a modified Wightman framework in which the Lorentz group is replaced by ${\rm O}(1,1)\times{\rm SO}(2)$ as the symmetry group. On the basis of this restricted symmetry, they also propose a modified locality condition (see  \cite{ChuFurutaInami:2006} for related perturbative calculations), which in our notation reads
\begin{align}\label{wwloc}
 [\phite_k(x),\phite_l(y)]_\pm=0\,,\qquad (x-y)\in W_1\cup(-W_1)
\,.
\end{align}
This vanishing of (anti-) commutators between field operators with the same $\te$ is however {\em not} a feature of models of the type considered here if $\te\neq 0$. Explicitly, one can for example consider the $\te$-deformed free scalar field $\phite_o$ and evaluate the two-particle contribution of the field commutator on the vacuum \cite{GrosseLechner:2007},
\begin{align*}
 \langle p_1,p_2\,|\, [\phi^{\te_1}_o(x), \phi^{\te_1}_o(y)]\,\Om\rangle
&=
-2i\left(
e^{i(p_1x+p_2y)}\,
-
e^{i(p_2x+p_1y)}
\right)
\sin\frac{p_1\te_1 p_2}{2}\,.
\end{align*}
For generic on-shell momenta $p_1, p_2$, this distribution does not vanish if $x-y$ are wedge-like separated as in \eqref{wwloc}. In fact, the interplay between $\phite$ and $\phi^{-\te}$ is essential to derive wedge locality, as was demonstrated in the proof of Theorem \ref{thm:wedge-locality}.
\\
\\
Regarding the discussion of the Spin-Statistics Theorem \cite{StreaterWightman:1964} for quantum field theories on noncommutative Minkowski space, we mention that this structure is undeformed in our framework: As is apparent from Lemma \ref{lemma-trans}, a deformed field $\phite_k$ transforms under a half-integer or integer spin representation precisely if its undeformed counterpart $\phi_k$ does. Also the modified commutation relations fit into this picture: Deformed fields $\phite_k(x), \phi^{\te'}_l(y)$ commute or anticommute precisely if $\phi_k(x), \phi_l(y)$ do, with the only modification that the condition of spacelike separation now also involves the parameters $\te,\te'$ and their associated wedge regions.
\\\\
The wedge-locality of the $\te$-deformed fields is of conceptual interest, since it shows that some restricted form of locality is still present also in the noncommutative setting. On the other hand, such localization properties are also useful from a more practical point of view since they allow for the computation of noncommutative corrections to the two-particle S-matrix.

If the initial undeformed theory has a decent energy-momentum spectrum, there exist two-particle incoming and outgoing scattering states $|p,q\rangle_{\rm in/out}^\te$, $\te\in\Sigma_{\kae\kam}$, also in the wedge-local deformed theory. Such asymptotic states have been constructed in \cite{BuchholzSummers:2008} using methods developed in \cite{BorchersBuchholzSchroer:2001}, the main ingredient being the fact that the two wedges $W(\te)$ and $W(\te')=W(-\te)$ can be causally separated. For a computation in the model of the $\te$-deformed free field, see \cite{GrosseLechner:2007}.

These scattering states can be used to calculate the S-matrix elements for collision processes with two incoming and two outgoing particles. Assuming for simplicity that the original theory describes a single species of massive particles, consider on-shell momenta $p,q,p',q'$ such that $q-p\in W(\te)$, $q'-p'\in W(\te)$. The S-matrix elements of the corresponding asymptotic two-particle states of the deformed theory are then given by \cite{BuchholzSummers:2008}
\begin{align}\label{scat-states}
 {}^{\,\;\;\; \te}_{\rm out}\langle p,q\,|\,p',q'\rangle_{\rm in}^\te
=
e^{-\frac{i}{2}\,p\te q} e^{-\frac{i}{2}\,p'\te q'}\,
\cdot
 {}^{\,\;\;\; 0}_{\rm out}\langle p,q\,|\,p',q'\rangle_{\rm in}^0
\,,
\end{align}
where ${}^{\,\;\;\; 0}_{\rm out}\langle p,q\,|\,p',q'\rangle_{\rm in}^0$ denote the S-matrix elements of the undeformed, local theory at $\te=0$. 

This deformation of the S-matrix shows that the effective interaction between particles on noncommutative Minkowski space is influenced by the noncommutativity. To detect this effect, one could for example use time delay experiments.

The asymptotic states \eqref{scat-states} depend on the noncommutativity parameter $\te$ of the fields $\phite$, $\phi^{-\te}$ used for preparing them. As pointed out in \cite{BuchholzSummers:2008}, the ordering of momenta with respect to the wedge $W(\te)$ breaks the Lorentz invariance of the S-matrix, a striking consequence of the nonlocality of the deformed models considered here.

\section{Conclusions}\label{sec:conclusion}

In the context of quantum field theory on noncommutative spacetimes, two quantization steps are involved: The usual quantization relating a classical field theory to a quantum field theory, and in addition the step from a ``classical spacetime'' to a noncommutative ``quantum spacetime''. Introducing a parameter $\vartheta$ measuring the noncommutativity of the spacetime, with $\vartheta=0$ corresponding to a commutative manifold\footnote{For the noncommutative Minkowski space considered here, one can use $\vartheta:=|\kae|+|\kam|$.}, the challenge is to formulate models of quantum matter on quantum spacetime, i.e. at $\hbar>0$ and $\vartheta>0$.

Starting from a classical field theory (CFT) on a classical spacetime (CST), at least two quite different strategies of constructing QFT on quantum spacetime (QST) are conceivable.
\begin{diagram}
\begin{array}{cc}\text{CFT on QST}\\(\hbar=0,\vartheta>0)\end{array} & \rTo^{\qquad\qquad}  & \begin{array}{cc}\text{QFT on QST}\\(\hbar>0,\vartheta>0)\end{array}\\
\uTo & & \uTo \\
\begin{array}{cc}\text{CFT on CST}\\(\hbar=0,\vartheta=0)\end{array} & \rTo & \begin{array}{cc}\text{QFT on CST}\\(\hbar>0,\vartheta=0)\end{array}
\end{diagram}
In the above diagram, one possible strategy consists in first formulating a model of classical fields on quantum spacetime. In concrete examples, this is usually done by considering deformed classical Lagrangeans, involving Moyal-products like $\varphi(x)\star...\star\varphi(x)$ as interaction terms. The second step in this procedure then consists in going over to a {\em quantum} field theory on QST, and is usually approached by perturbative renormalization with new, $\vartheta$-dependent counter terms to define the corresponding QFT on QST (see, for example, \cite{GrosseWulkenhaar:2005, GurauMagnenRivasseauTanasa:2008, GrosseVignesTourneret:2008} for Euclidean models, and \cite{BahnsDoplicherFredenhagenPiacitelli:2005} for a Lorentzian approach).

A different strategy consists in taking the other route from the lower left corner at $(\hbar=0,\vartheta=0)$ to the upper right corner at $(\hbar>0,\vartheta>0)$ in the  above diagram. This amounts to first considering a QFT on a classical spacetime, and then applying the deformation to quantum spacetime afterwards \cite{AlvarezGaumeVazquezMozo:2003, GrosseLechner:2007, Soloviev:2008, BuchholzSummers:2008}.

These two alternative procedures are however inequivalent in general, i.e. the described diagram is not commutative. This can for example be seen when considering the theory of a free, scalar field on Moyal space: Arguing that the corresponding classical Lagrangean $L_\vartheta$ on QST should arise from the initial free Lagrangean $L_0$ by replacing ordinary products with Moyal $\star$-products, one sees that the action is unchanged since $L_\vartheta$ is quadratic (see, e.g. \cite{Rivasseau:2007}). Hence ``noncommutative free QFTs'' are undeformed from this point of view, i.e. identical with the usual free QFTs on commutative spaces. However, following the second strategy, one arrives at the conclusion that the deformed free theory does differ from its commutative counterpart \cite{GrosseLechner:2007}. 
\\\\
In the present paper, we discussed an approach in the spirit of the second construction procedure. In the formalism presented here, the noncommutative structure of spacetime amounts to a universal deformation of the QFT under consideration, which can be traced back to a deformation of the tensor product in the underlying Borchers-Uhlmann algebra of test functions. The differences and similarities between the basic structures of a usual quantum field theory and a deformed one were investigated.

In view of the simple form the noncommutativity, it seems well possible to extend our formalism to other topics, such as thermal equilibrium states of deformed quantum field theories, or a Euclidean formulation and its relation to the Minkowski regime. Moreover, it would be interesting to understand better the relation between the two different construction strategies pointed out in the above diagram, and to analyze the interplay between the $\te$-deformation and perturbation theory.
\\\\
Independently of the motivation to study quantum field theory on noncommutative Minkows\-ki space, the construction carried out here is also of interested for usual ``commutative'' QFT, as emphasized in \cite{BuchholzSummers:2008}: It provides us with new wedge-local, covariant models with non-trivial S-matrix. Using methods of algebraic quantum field theory \cite{Haag:1996}, the {\em local} observable content of such models can be determined and used to define a strictly local theory. In the present $\te$-deformed setting, the corresponding local models are expected to be trivial \cite{BuchholzSummers:2008}, which is consistent with the nonlocal structure of Moyal space. But the general strategy of constructing local, interacting models from deformed wedge-local theories seems to be a promising new approach to constructive quantum field theory, which in the two-dimensional case has already led to the rigorous construction of many models which were not accessible by other methods \cite{Schroer:1997-1, Lechner:2003, BuchholzLechner:2004, BuchholzSummers:2007, Lechner:2008-1}.

In this context, we briefly mention possible generalizations of the deformation discussed in this paper. The $\te$-deformation amounts to equipping the Borchers-Uhlmann algebra underlying Wightman theory with a new ``twisted'' tensor product, the Moyal tensor product $\otte$. However, most of the structural results derived here do not depend on the specific form of this twisted tensor product, but rather hold for more general deformations. For example, one can define a product $\otte^\rho$ as
\begin{align}
\widetilde{(f^n\otte^\rho g^m)}(p,q)
:=
\prod_{l=1}^n\prod_{r=1}^m \tilde{\rho}(p_l\te q_r)\cdot \widetilde{f^n}(p)\widetilde{g^m}(q)
\,,
\end{align}
with a suitable function $\tilde{\rho}$ satisfying in particular $\tilde{\rho}(0)=1$ and $\tilde{\rho}(-\la)=\overline{\tilde{\rho}(\la)}$. For functions of a single variable, this new product arises by smearing over a range of noncommutativities,
\begin{align}
 (f^1\otte^\rho g^1)(x,y)
&=
\int_0^\infty \frac{d\la\,\rho(\la)}{\sqrt{2\pi}}\,(f^1\ot_{2\la\cdot\te} g^1)(x,y)\,.
\end{align}
Provided that the vacuum state on the Borchers-Uhlmann algebra is compatible with this new product in the sense that $\{f\in\uSs\,:\,\om(f^*\ot f)=0\}$ is a left ideal with respect to multiplication with $\otte^\rho$, the corresponding fields $\phi^{\te,\rho}(f)\Psi(g):=\Psi(f\otte^\rho g)$ are well defined. They then satisfy the same covariance (Lemma \ref{lemma-trans} and Proposition \ref{Prop:TCP}) and locality properties (Theorem \ref{thm:wedge-locality}) as in the previously considered case corresponding to $\rho(\la)=\sqrt{2\pi}\,\delta(\la-\frac{1}{2})$, and lead to a non-trivial S-matrix involving $\rho$. 

Constructions of this type therefore further illustrate the possibility of obtaining quantum field theories with various non-trivial S-matrices from interaction-free theories by means of a deformation procedure. This topic will be studied in more detail in a forthcoming publication\footnote{G.~Lechner, in preparation.}.

\acknowledgments
The hospitality and financial support of the Erwin Schr\"odinger Institute for Mathematical Physics in Vienna is gratefully acknowledged.

\bibliography{$HOME/Physik/Artikelsammlung/Bibtex-Database.bib}
\bibliographystyle{JHEP}
\end{document}